# Colossal optical anisotropy from atomic-scale modulations

Hongyan Mei[1†], Guodong Ren[2†], Boyang Zhao[3†], Jad Salman[1†], Gwan Yeong Jung[4], Huandong Chen[3], Shantanu Singh[3], Arashdeep S. Thind[2], John Cavin[5], Jordan A. Hachtel[6], Miaofang Chi[6], Shanyuan Niu[3], Graham Joe[1], Chenghao Wan[1,7], Nick Settineri[8], Simon J. Teat[8], Bryan C. Chakoumakos[9], Jayakanth Ravichandran[3,10,11*], Rohan Mishra[2,4*], Mikhail A. Kats[1,7*]

*[1] Department of Electrical and Computer Engineering, University of Wisconsin-Madison, Madison, WI 53706, USA*
*[2] Institute of Materials Science and Engineering, Washington University in St. Louis, St. Louis, MO 63130, USA*
*[3] Mork Family Department of Chemical Engineering and Materials Science, University of Southern California, Los Angeles, CA 90089, USA*
*[4] Department of Mechanical Engineering and Material Science, Washington University in St. Louis, St. Louis, MO 63130, USA*
*[5] Department of Physics, Washington University in St. Louis, St. Louis, MO 63130, USA*
*[6] Center for Nanophase Materials Sciences, Oak Ridge National Laboratory, Oak Ridge, TN 37831, USA*
*[7] Department of Materials Science and Engineering, University of Wisconsin-Madison, Madison, WI 53706, USA*
*[8] Advanced Light Source, Lawrence Berkeley National Laboratory, Berkeley, CA 94720, USA*
*[9] Neutron Scattering Division, Oak Ridge National Laboratory, Oak Ridge TN 37831, USA*
*[10] Ming Hsieh Department of Electrical Engineering, University of Southern California, Los Angeles, CA 90089, USA*
*[11] Core Center for Excellence in NanoImaging, University of Southern California, Los Angeles, CA 90089, USA*
*[†] These authors contributed equally to this work*
**Email: mkats@wisc.edu, rmishra@wustl.edu, jayakanr@usc.edu*

## Summary paragraph

In modern optics, materials with large birefringence ($\Delta n$, where $n$ is the refractive index) are sought after for polarization control (*e.g.* in wave plates, polarizing beam splitters, etc.[1–3]), nonlinear optics and quantum optics (*e.g.* for phase matching[4,5] and production of entangled photons[6]), micromanipulation[7], and as a platform for unconventional light-matter coupling, such as Dyakonov-like surface polaritons[8] and hyperbolic phonon polaritons[9–11]. Layered "van der Waals" materials, with strong intra-layer bonding and weak inter-layer bonding, can feature some of the largest optical anisotropy[12–16]; however, their use in most optical systems is limited because their optic axis is out of the plane of the layers and the layers are weakly attached, making the anisotropy hard to access. Here, we demonstrate that a bulk crystal with subtle periodic modulations in its structure — $Sr_{9/8}TiS_3$ — is transparent and positive-uniaxial, with extraordinary index $n_e = 4.5$ and ordinary index $n_o = 2.4$ in the mid- to far-infrared. The excess Sr, compared to stoichiometric $SrTiS_3$, results in the formation of $TiS_6$ trigonal-prismatic units that break the infinite chains of face-shared $TiS_6$ octahedra in $SrTiS_3$ into periodic blocks of five $TiS_6$ octahedral units. The additional electrons introduced by the excess Sr subsequently occupy the $TiS_6$ octahedral blocks to form highly oriented and polarizable electron clouds, which selectively boost the extraordinary index $n_e$ and result in record birefringence ($\Delta n > 2.1$ with low loss). The connection between subtle structural modulations and large changes in refractive index suggests new categories of anisotropic materials and also tunable optical materials with large refractive-index modulation and low optical losses.

## Main text

Birefringence is the dependence of the refractive index on the polarization of light travelling through a material. The observation of birefringence in calcite as early as 1669[17]—called Iceland spar at the time—eventually led to Fresnel's insight in 1821 that light is a transverse wave[18,19]. Calcite's record as the most birefringent material stood for over a century, with $\Delta n = |n_e - n_o| = 0.17$ in the visible, as analyzed and explained by Bragg[20]; here, $n_e$ and $n_o$ are respectively the extraordinary and ordinary refractive index. In calcite ($CaCO_3$) and other calcite-type carbonates ($RCO_3$; $R$ = Mg, Zn, Fe, Mn, and others), the anisotropy primarily results from the interaction of dipole excitations around the oxygen atoms within the planar carbonate ions ($CO_3^{2-}$), which are all oriented perpendicular to the optic axis within the crystal[20–23]. Achieving much larger optical birefringence is expected to require much larger structural anisotropy.

Indeed, the revolution of layered (two-dimensional or 2D) materials has led to the demonstration of many crystals with very large optical anisotropy due to strong intra-layer bonding (covalent or ionic) and weak inter-layer bonding (van der Waals), resulting in, e.g., $\Delta n \sim 0.7$ in hexagonal boron nitride (h-BN)[13] in the visible and near infrared and $\Delta n \sim 1.5$ in molybdenum disulfide ($MoS_2$)[16] in the near infrared [Fig. 1(e, f)]. However, the giant anisotropy found in (usually thin) layered crystals is difficult to exploit for either bulk optics or micro-optics because their optic axis is out of the plane of the layers and the layers are weakly bonded. Therefore, there is a need to discover or engineer bulk materials with giant anisotropy, especially in the infrared. In 2018, our groups[i] reported that $BaTiS_3$, a quasi-one-dimensional (quasi-1D) hexagonal perovskite chalcogenide with face-shared ($TiS_6$) octahedral chains, has $\Delta n = 0.76$ at mid-infrared frequencies where it is transparent—a record at the time[24,25]. These quasi-1D hexagonal chalcogenide single crystals[26] and thin films[27] can be grown with different orientations to enable easy access to their anisotropic properties. Therefore, they are an attractive and largely unexplored class of materials to achieve higher birefringence and dichroism. At the time, we selected Ba, S, and Ti ions and studied $BaTiS_3$ due to the large contrast in the electronic polarizability of the individual candidate ions in the $ABX_3$ quasi-1D structure, and thus believed that the birefringence we demonstrated is close to the limit for this class of materials[24].

Here, we introduce structural modulation as a new mechanism for dramatically enhancing the anisotropy of electronic polarizability, far exceeding values that have been achievable by the anisotropic distribution of individual ions with distinct polarizability. In quasi-1D chalcogenide $Sr_{1+x}TiS_3$, the structural modulation controls the selective occupation of strongly oriented (anisotropic) electronic states, and hence leads to a birefringence of ~ 2.1, significantly larger than has been observed in transparent regions of both quasi-1D and layered "van der Waals" materials to date [Fig. 1(e, f)].

[i] Only some authors overlap between the two papers

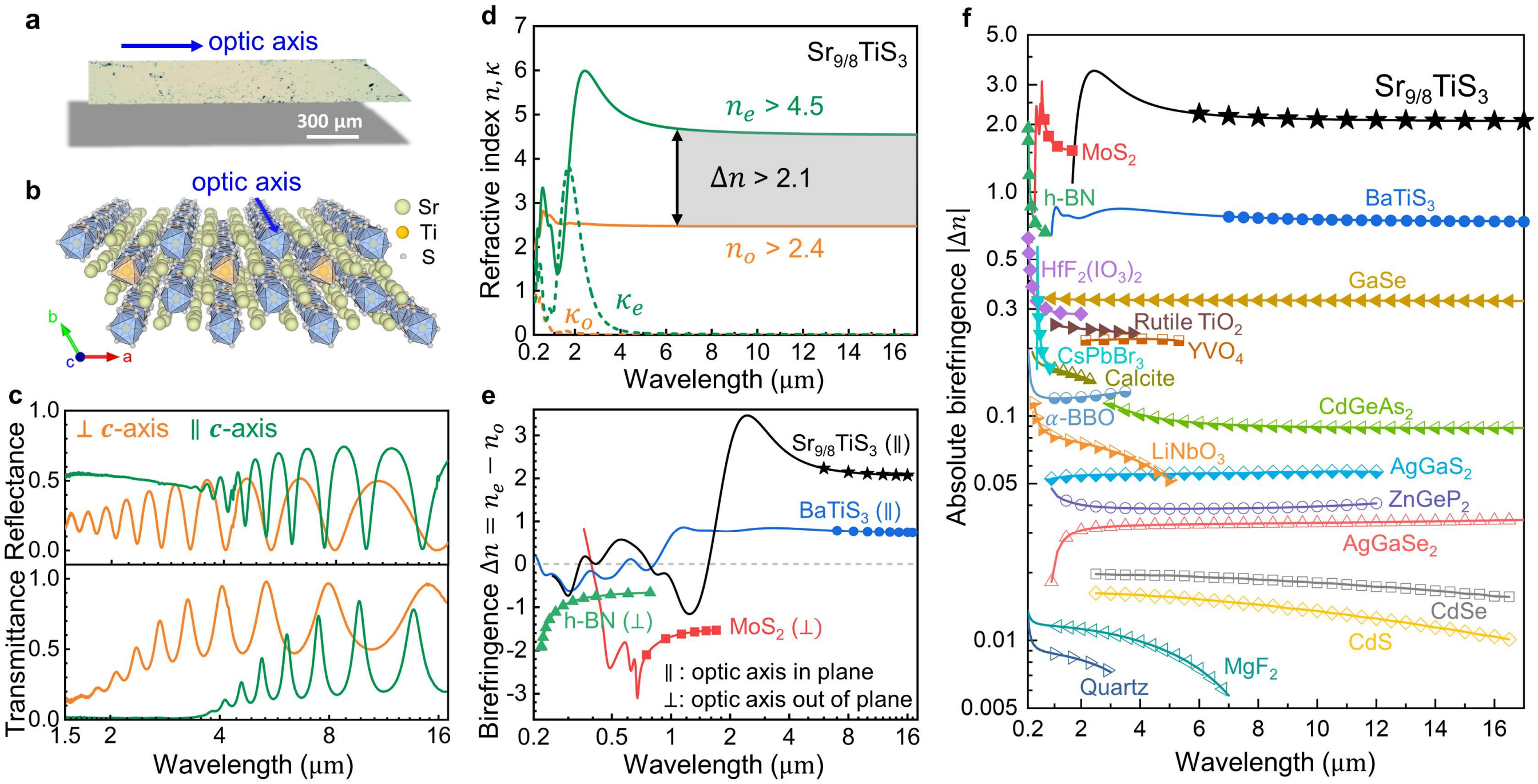

**Figure 1. Infrared birefringence $\Delta n > 2.1$ in $Sr_{9/8}TiS_3$ single crystals. a**. Optical image of the $Sr_{9/8}TiS_3$ crystal measured in this work, with scale bar. **b**. Perspective schematic of a $Sr_{9/8}TiS_3$ crystal with the optic axis along the $TiS_6$ chains. **c**. Polarized reflectance and transmittance of a $Sr_{9/8}TiS_3$ crystal plate with thickness = 3.2 μm measured across the mid infrared (the wavelength axis uses the log scale). **d**. Extracted complex refractive-index values of uniaxial $Sr_{9/8}TiS_3$ for the ordinary (perpendicular to *c*-axis, orange) and extraordinary (parallel to *c*-axis, green) directions spanning the visible through the mid infrared, based on a combination of spectroscopic ellipsometry and transmittance/reflectance measurements. For wavelengths longer than ~6 μm, the material is highly transparent and maintains a large birefringence, $\Delta n$ > 2.1 (gray shaded region). **e.** Comparison of the birefringence of two representative hexagonal quasi-1D chalcogenides ($Sr_{9/8}TiS_3$ and $BaTiS_3$[24]) with highly anisotropic 2D materials (h-BN[13] and $MoS_2$[16]). The transparent regions of these materials are identified using symbols. **f.** Comparison of the absolute birefringence values of $Sr_{9/8}TiS_3$ and a variety of anisotropic materials [from the literature[13,16,24,28–37]], showing that $Sr_{9/8}TiS_3$ has by far the largest birefringence among reported anisotropic crystals. The symbols indicate regions of transparency.

## Synthesis and basic structure of strontium titanium sulfide crystals

We synthesized single crystals of $Sr_{9/8}TiS_3$ using the chemical vapor transport method, which has been reported earlier[24,26,38,39]. $Sr_{1+x}TiS_3$ falls in a broad category of $BaNiO_3$-related structures, with chemical formula $A_{1+x}BX_3$ ($A$ = alkaline metal, $B$ = transition metal, $X$ = anion)[40,41]. It has quasi-1D chains of face-shared $BX_6$ octahedra that are aligned along a 6-fold rotational axis (commonly the *c*-axis) with $A$ cations filling the inter-chain interstitials. Its chemical composition was determined to be off-stoichiometric using energy dispersive analytical X-ray spectroscopy (EDS) as reported in a previous work[39]. Although $A_{1+x}BX_3$ compounds are commonly stoichiometric (i.e., $x = 0$), certain non-stoichiometric crystalline structures, such as $Sr_{9/8}TiS_3$ and $Sr_{8/7}TiS_3$, have been reported to be more thermodynamically stable[42,43]; in these structures, excess Sr atoms periodically compress the Sr-lattice while introducing a stacking sequence of distorted $TiS_6$ polyhedral and displaced Sr atoms along the *c*-axis, expressed as structural modulation[40,42,43].

## Measurement of colossal birefringence

The optical anisotropy of $Sr_{9/8}TiS_3$ was measured in two steps: we first acquired polarization-dependent, normal-incidence reflectance and transmittance spectra of the plate in Fig. 1(a) using Fourier-transform infrared spectroscopy (FTS), shown in Fig. 1(c). The thickness of the crystal was estimated to be 3.2 μm by fitting to the Fabry–Pérot fringes in the spectra for each polarization, and verified by cross-section scanning electron microscope (SEM) imaging (Fig. S3 in *Supplementary Section 2*). The large difference in the reflectance and transmittance between the two polarizations (parallel and perpendicular to the *c*-axis) is a clear indication of large optical anisotropy. For wavelengths longer than ~6 μm, we observe low optical losses up to the cut-off wavelength of the detector at 17 μm [Fig. 1(d)]. We expect that this low-loss region extends to longer wavelengths, limited by phonon (~27 μm)[39] or plasmon resonance (~100 μm, calculated in *Supplementary Section 2*).

To fully quantify the degree of optical anisotropy, we then combined variable-angle ellipsometry measurements over the spectral range of 210 nm to 2500 nm with the polarization-resolved reflection and transmission measurements in Fig. 1(c), and extracted the complex refractive index of $Sr_{9/8}TiS_3$ for wavelengths from 210 nm to 17 μm [Fig. 1(d)]. Three different ellipsometry measurement were performed: with the c-axis parallel to the plane of incidence, perpendicular to the plane of incidence, and at an off-axis angle; all of this data was combined in a simultaneous analysis (*see more details in Supplementary Section 2*). In the low-loss region of $\lambda > 6$ μm, $Sr_{9/8}TiS_3$ has a birefringence ($\Delta n$) up to 2.1. This is by far the largest birefringence among reported anisotropic crystals, to the best of our knowledge [Fig. 1(f)]. Compared to layered 2D materials $MoS_2$[16] and h-BN[13], $Sr_{9/8}TiS_3$ possesses large anisotropy across a broad low-loss region in mid-infrared [Fig. 1(e)], and the in-plane optic axis is easy to exploit for practical optical components. As we discuss below, this extreme birefringence is a result of the enhancement of the extraordinary index ($n_e$) from the expected ≲3.4 in a hypothetical unmodulated stoichiometric crystal to ~4.5 in $Sr_{9/8}TiS_3$.

## Direct observation of structural modulations in $Sr_{9/8}TiS_3$ crystals

The crystal structure of $Sr_{9/8}TiS_3$ was resolved using single-crystal X-ray diffraction (SC-XRD). Both 3D and (3+1)D modulation approaches were used to solve the modulated structure (*Supplementary Section 1*). The resulting $Sr_{54}Ti_{48}S_{144}$ structure of *R*3*c* and $Sr_{1.125}TiS_3$ structure of $R\bar{3}m(00g)0s$[44] space groups reveal a similar modulated structure (visualized in Fig. 2(a, b)), which is consistent with the previously reported *R*3*m*(00g)0*s* $Sr_{9/8}TiS_3$ structure[42,43,45–47]. The details of the data collection, data reduction, and structure refinements are listed in Tables S1 in *Supplementary Section 1*, and the best resulting crystal structure is reported in Table S2 and Table S3 in *Supplementary Section 1*.

In contrast to the hypothetical stoichiometric counterpart $SrTiS_3$ (*Supplementary Section 3* Fig. S8), the $Sr_{9/8}TiS_3$ lattice has structural modulations consisting of blocks of face-shared octahedra (referred as 'O') that are separated by pseudo-trigonal-prismatic $TiS_6$ units (referred as 'T') along the *c*-axis as shown in Fig. 2(b). The structural modulation of $Sr_{9/8}TiS_3$ arises from an overall trigonal twist distortion compared to the average unmodulated structure of $SrTiS_3$. To accommodate excess Sr in the lattice along the *c*-axis, Sr atoms undergo displacements within the *ab* plane, resulting in the triangular-shaped projection in Fig. 2(a). The Sr displacements are accompanied by twist distortion of $TiS_6$ units from octahedral to trigonal-prismatic polyhedra. These different polyhedral units have different Ti–Ti distances along the *c*-axis (Fig. S9 and S10 in *Supplementary Section 5*). By counting the stacking sequence of building blocks (structurally classified as O and T), $TiS_6$ chains with periodic $[-(T-O-T)-(O)_5-]_2$ succession can be used to define the modulation periodicity of 16 units of $TiS_6$ within every 18 Sr layers in the $Sr_{9/8}TiS_3$ lattice [Fig. 2(b)].

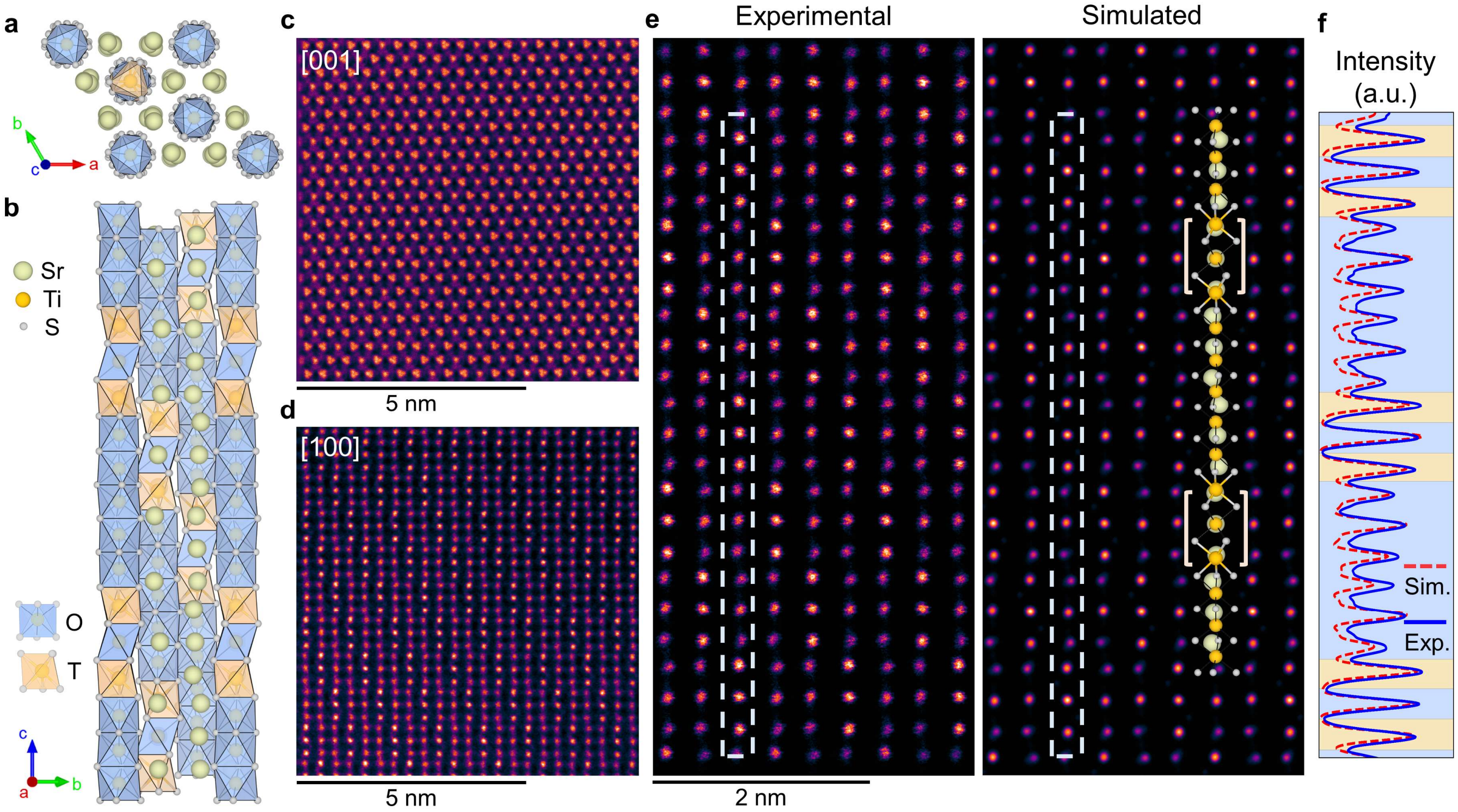

**Figure 2. Structural modulation in $Sr_{9/8}TiS_3$. a, b.** Schematics representing the modulated $Sr_{9/8}TiS_3$ (*R*3*c*) lattice, resolved from single-crystal XRD (SC-XRD), and viewed along the [001] axis in **a**, and [100] axis in **b**. The octahedral (O) and pseudo-trigonal prismatic (T) $TiS_6$ units are highlighted in blue and orange, respectively. **c, d.** Atomic-resolution HAADF-STEM images of a $Sr_{9/8}TiS_3$ crystal along the [001] axis in **c**, and [100] axis in **d**. **e.** High-magnification HAADF-STEM image (left panel) and simulated image (right panel) of $Sr_{9/8}TiS_3$ view along the [100] axis. A schematic of one column of atoms is overlayed on the simulated image. A repeating pattern of three bright atomic columns, where the Ti and Sr atoms overlap along the viewing direction, can be observed in these HAADF images. These triplet atomic columns are highlighted with square brackets in the atomic model. **f.** Line profiles across the experimental and simulated STEM images (white boxes in **e**) comparing the intensity variation across a single atomic column.

To directly visualize the subtle structural modulations in $Sr_{9/8}TiS_3$, we performed atomically resolved imaging using an aberration-corrected scanning transmission electron microscope (STEM). Large field-of-view, high-angle annular dark-field (HAADF) images of the $Sr_{9/8}TiS_3$ crystal viewed along the [001] and [100] zone axis are shown in Fig. 2(c, d). In this imaging mode, the intensity of the atomic columns is approximately proportional to the square of the effective atomic number of the column ($Z^2$)[48]. Along the [001] zone axis, the Sr atomic columns appear as triangles due to their staggered arrangement along the *c*-axis [Fig. 2(c)], which match well with the structural features in Fig. 2(a). Along the [100] orientation, the Ti and Sr columns overlap within the triple blocks of – (T–O–T) –, and therefore, they appear as bright triplets in the HAADF images as they have higher intensity than the Sr-only atomic columns within the block of five octahedral units $-(O)_5-$ [Fig. 2(d, e)]. We also observe periodic distortions of the Sr atomic columns in the form of contraction and dilation of the Sr–Sr distance between neighboring chains (see more discussions in *Supplementary Section 5*). A comparison of the intensity and spacing between atomic columns in the experimental and the simulated HAADF images, as shown in Fig. 2(f), shows excellent agreement, and corroborates the modulation periodicity.

## Electronic structure and calculation of dielectric properties

To reveal the origin of the giant optical anisotropy in $Sr_{9/8}TiS_3$ and its relationship with structural modulations, we performed first-principles density-functional theory (DFT) calculations. To understand the formation of the modulated $Sr_{9/8}TiS_3$ phase instead of $SrTiS_3$, we performed a convex hull analysis using the DFT calculated energies of the two compounds and all possible lower-order decomposition products (*Supplementary Section 4* Table S5). We find that modulated $Sr_{9/8}TiS_3$ is on the hull, and is thus stable against decomposition, as opposed to stoichiometric $SrTiS_3$, which is thermodynamically metastable, being 45 meV/atom above the hull.

Next, we calculated the electronic structures of $Sr_{9/8}TiS_3$ and $SrTiS_3$ to understand the effect of modulations on the optical properties. Both $SrTiS_3$ and $Sr_{9/8}TiS_3$ are computed to possess an indirect bandgap. $SrTiS_3$ has an indirect bandgap between valence band maximum (VBM) at Γ point and conduction band minimum (CBM) at A point. The topmost valence band and bottom of the conduction band of $SrTiS_3$ show a relatively flat behavior along all the paths in the Brillouin zone, except for the Γ−Z direction, which in the reciprocal space corresponds to the direction parallel to the *c*-axis where the neighboring $TiS_6$ octahedra have face-sharing connectivity, while octahedral connectivity is broken along the *ab*-plane[49,50]. Compared with $SrTiS_3$, $Sr_{9/8}TiS_3$ shows similarly flat bands. The topmost valence bands and bottom conduction bands arise from *d*-states, and form the indirect bandgap between VBM at Γ and CBM at T.

In $Sr_{9/8}TiS_3$, the electrons introduced by excess $Sr^{2+}$ cations occupy the nominally empty Ti *d* states. Using DFT + Hubbard *U* calculations[51], with a $U = 3.0$ eV for the Ti atoms, we checked for different magnetic orderings of the moments and found the paramagnetic configuration to be the most stable. The details of the calculations can be found in *Methods* and *Supplementary Section 8*. As shown in the calculated band structure of $Sr_{9/8}TiS_3$ in Fig. 3(b), the additional valence electrons preferentially occupy $3d_{z^2}$ states (highlighted in red) of Ti atoms in $Sr_{9/8}TiS_3$. These selectively occupied energy states in modulated $Sr_{9/8}TiS_3$ can be corroborated by atomic-resolution electron energy-loss spectroscopy (EELS), which exhibits subtle but distinct differences between the pseudo-trigonal prismatic $TiS_6$ units (T) and the octahedral $TiS_6$ units (O) (*Supplementary Section 7* Fig. S14). Compared to the pseudo-trigonal prismatic $TiS_6$ units (T), octahedral $TiS_6$ units (O) have shorter Ti-Ti distance, resulting in $3d_{z^2}$ states that are lower in energy. Thus, the Ti atoms in the O block preferentially accept the additional electrons (*Supplementary Section 6* Fig. S13). This also opens up a band gap between the occupied $3d_{z^2}$ states of the octahedrally coordinated Ti atoms and the unoccupied Ti-$3d_{z^2}$ states of the Ti atoms with trigonal-prismatic coordination. The character of the edge states in modulated $Sr_{9/8}TiS_3$ is in sharp contrast to that of $SrTiS_3$ wherein the band gap is between the S-3$p$ states in the valence band and Ti-$3d_{z^2}$ states in the conduction band [Fig. 3(a)].

With the electronic ground states computed, we then calculated the complex dielectric function $\varepsilon_{\parallel/\perp}(\omega) = \varepsilon_{1\parallel/\perp}(\omega) + i\varepsilon_{2\parallel/\perp}(\omega)$ for electric fields along (||) and perpendicular to (⊥) the *c*-axis. The imaginary part $\varepsilon_2(\omega)$ is obtained by calculating the direct transitions between occupied and unoccupied states[52]. The real part $\varepsilon_1(\omega)$ is then extracted by a Kramers-Kronig transformation (see details in *Supplementary Section 8*). Fig. 3(e-f) show the frequency-dependent dielectric functions of the stoichiometric $SrTiS_3$ and the modulated $Sr_{9/8}TiS_3$, with very similar results perpendicular to the *c*-axis ($\varepsilon_{1\perp}$), but dramatic enhancement of the dielectric function parallel to the *c*-axis ($\varepsilon_{1\parallel}$). The enhancement is a consequence of the selective occupation of $d_{z^2}$ states at the $(O)_5$ segments in modulated $Sr_{9/8}TiS_3$, which we show in real space by using an isosurface plot of the charge density arising from the occupied $3d_{z^2}$ band [Fig. 3(d)]. The occupied $3d_{z^2}$ electrons form a highly oriented blob and result in additional polarizability along the optic axis ($\varepsilon_{1\parallel}$). In contrast, the electrons from the valence band states in stoichiometric $SrTiS_3$ have an isotropic character and are localized on the S atoms [Fig. 3(c)]. The unoccupied conduction band is at substantially higher energy compared to the $3d_{z^2}$ electrons, resulting in very few free carriers and therefore low free-carrier absorption.

While an experimental comparison between $Sr_{9/8}TiS_3$ and $SrTiS_3$ cannot be made due to the metastable nature of $SrTiS_3$, we did compare the optical properties of $Sr_{9/8}TiS_3$ to $BaTiS_3$, and to hypothetical $SrTiS_3$ which is isostructural to $BaTiS_3$ in Fig. S19 in *Supplementary Section 8.*

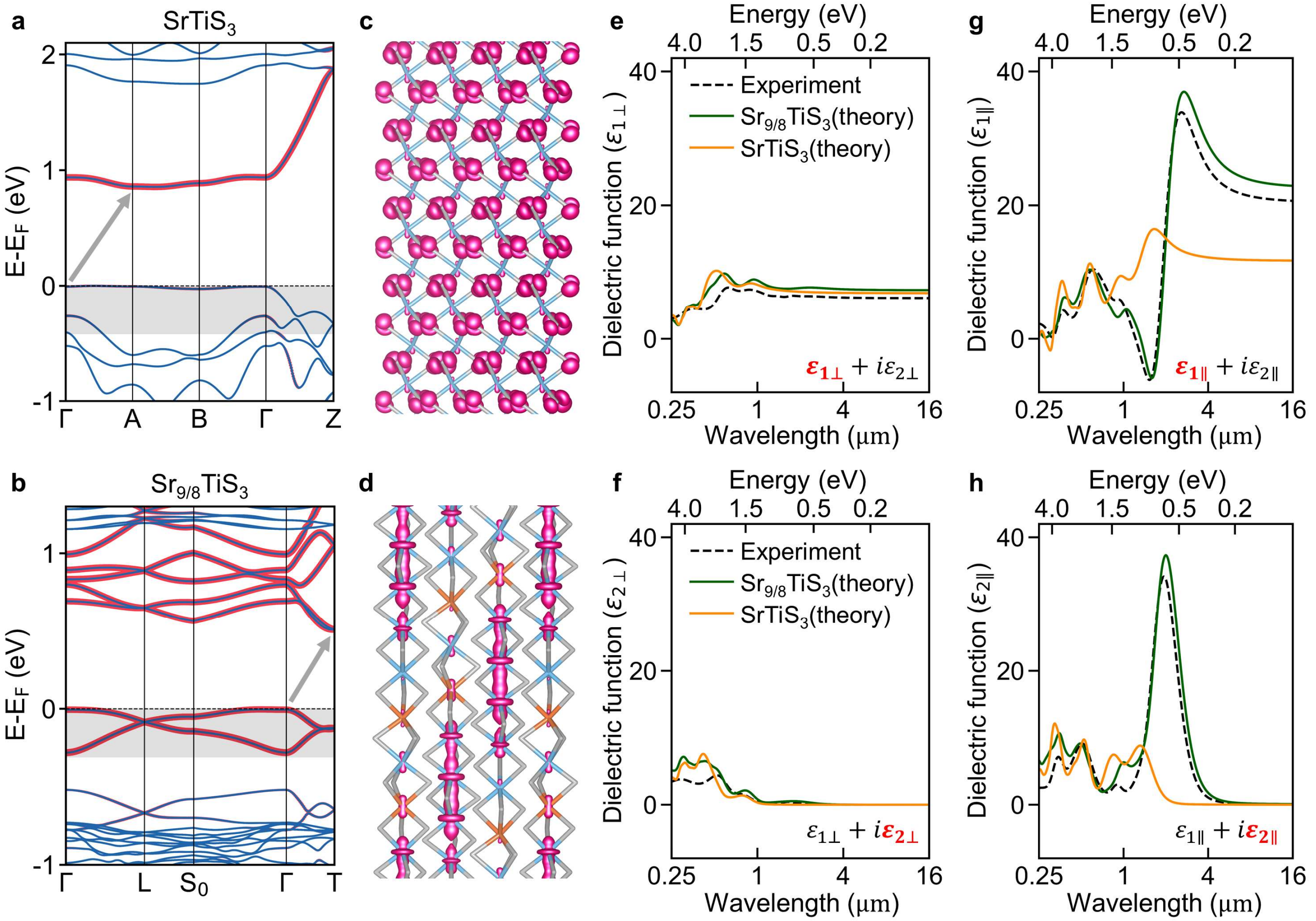

**Figure 3. Electronic structure and optical properties of modulated $Sr_{9/8}TiS_3$. a-b.** Orbital-projected band structures for hypothetical stoichiometric $SrTiS_3$ in **a**, and modulated $Sr_{9/8}TiS_3$ in **b**. The thicker lines highlighted in red correspond to the contribution from Ti-3$d_{z^2}$ states. The Fermi energy is set to 0 eV. **c-d**. Spatial distribution of the valence electrons below the Fermi energy (shaded in gray in a-b), showing **c.** S-3*p* character in $SrTiS_3$ and **d.** Ti-3$d_{z^2}$ character in $Sr_{9/8}TiS_3$. The isosurface is set to an electron density of 0.004 e/Å$^3$). **e-h**. Calculated complex dielectric function for polarization perpendicular ($\varepsilon_{i\perp}$) and parallel ($\varepsilon_{i||}$) to the *c*-axis of the hypothetical stoichiometric $SrTiS_3$ and modulated $Sr_{9/8}TiS_3$, compared to the experimental results (black dashed line). The index $i$ = 1 represents the real part of the dielectric function and 2 represents the imaginary part.

## Conclusion

In this article, we demonstrate how subtle atomic-scale structural modulations of a bulk crystal can dramatically change its optical properties. We synthesized and studied single-crystal plates of modulated quasi-1D chalcogenide $Sr_{9/8}TiS_3$, a uniaxial material which we found to possess a record birefringence ($\Delta n > 2.1$) in a broadband low-loss spectral region between $\lambda = 6$ μm and at least 17 μm, with the optic axis in plane of the samples. Compared to stoichiometric unmodulated $SrTiS_3$, which is expected to have $\Delta n \lesssim 1$, the extra Sr in $Sr_{9/8}TiS_3$ results in additional electrons that selectively occupy localized anisotropic states (Ti-3$d_{z^2}$), greatly enhancing the polarizability of the material along the optic axis, and thus resulting in a degree of optical anisotropy far larger than has been demonstrated in any bulk material. The atomic-scale structural features of $Sr_{9/8}TiS_3$ were resolved using single-crystal X-ray diffraction and directly observed with HAADF-STEM imaging, and the resulting structural information was used to perform DFT calculations that clarified the physical mechanism leading to the experimentally observed colossal optical anisotropy. We anticipate that structural modulation in nonstoichiometric crystals will be a new tool in realizing materials with large degrees of optical and optoelectronic anisotropy. Furthermore, the connection between subtle structural modulations and large changes in the refractive index may enable a new class of optical materials that can be tuned with an applied stimulus.

## Methods

### Crystal growth method

Single crystals of $Sr_{9/8}TiS_3$ were grown via the chemical vapor transport method with iodine as the transporting agent. Starting materials, strontium sulfide powder (Alfa Aesar, 99.9%), titanium powder (Alfa Aesar, 99.9%), sulfur pieces (Alfa Aesar, 99.999%) and iodine pieces (Alfa Aesar 99.99%) were stored and handled in a nitrogen-filled glove box. Stoichiometric quantities (weighed as SrS : Ti : S = 1 : 1 : 2) of precursor powders with a total weight of 1.0 g were mixed and loaded into a quartz tube (19 mm diameter and 2 mm thickness) along with ~0.75 mg·cm$^{-3}$ iodine inside the glove box. The tube was capped with ultra-torr fittings and a quarter-turn plug valve to avoid exposure to air before being evacuated and sealed using a blowtorch. The sealed tube was then loaded in an MTI OTF-1200X-S-II Dual Heating Zone 1200 °C compact split tube furnace, heated to the reaction temperature of 1055 °C at 100 °C/h, and held for 150 hours before cooling down. The temperature gradient in the dual-zone furnace was kept at 5 °C/cm.

### Infrared reflection and transmission spectroscopy

Polarization-resolved infrared spectroscopy was carried out using a Fourier transfer infrared spectrometer (Bruker Vertex 70) outfitted with an infrared microscope (Hyperion 2000). A 15× Cassegrain microscope objective (NA = 0.4) was used for both reflection and transmission measurements under normal incidence on the (100) face of a $Sr_{9/8}TiS_3$ crystal. These measurements were performed with a Globar source, a potassium bromide (KBr) beam splitter, and a mercury cadmium telluride (MCT) detector. A wire-grid polarizer was used to control the polarization of the incident light. The samples were maintained at room temperature.

### Spectroscopic ellipsometry

Variable-angle spectroscopic ellipsometry measurements were performed using a VASE ellipsometer with focusing probes (J. A. Woollam Co.) over a spectral range of 210 nm to 2500 nm at an angle of incidence of 55°. Data were acquired from three different sample orientations (optical axis parallel, perpendicular, and 30° to the plane of incidence). Data analysis and refractive index extraction were performed using WVASE software (J. A. Woollam Co.). When creating the optical model, we assumed that the crystal is uniaxial with the *c*-axis along the quasi-1D chains, as implied by our structural characterization in Fig. 2. The samples were maintained at room temperature. More details can be found in *Supplementary Section 2*.

**Single-crystal diffraction**

Single-crystal diffraction of $Sr_{9/8}TiS_3$ crystals was first collected using the Rigaku XtaLAB AFC12 (RCD3) diffractometer at Oak Ridge National Laboratory. The diffractometer is equipped with a Mo Kα X-ray source (wavelength 0.71073 Å) and a Rigaku HyPix-6000HE detector. Diffraction data were collected using four $\omega$ scans with a step size of 0.5° and a collection time of 25 seconds per step, which achieved a 99.95% completeness (total number of measured peaks divided by the total number of peaks), for a resolution of 0.75 Å.

A larger-size platelet-shaped crystal was characterized in beamline 12.2.1 at the Advanced Light Source (ALS) at Lawrence Berkeley National Laboratory. Crystals were mounted on MiTeGen Dual Thickness MicroMounts™ and placed in a nitrogen cold stream on the goniometer-head of a Bruker D8 diffractometer, which is equipped with a PHOTONII CPAD detector operating in shutterless mode. Diffraction data were collected using synchrotron radiation at a wavelength of 0.72880 Å with silicon (111) monochromator. A combination of $\varphi$ and $\omega$ scans with scan speeds of 1 s per 2 degrees for the $\varphi$ scans, and 1 s per 0.15 degree for the $\omega$ scans at $2\theta = 0$ and -20°, respectively, were acquired. 99.95% completeness for a resolution of 0.6 Å was achieved.

Data reduction, scaling and precession map analysis were done in CrysAlisPro and APEX3 corresponding to the original data collection format. Crystal structures were solved and refined in ShelXle[53] and Jana 2020[54].

**Electron microscopy**

We prepared [001] a cross-sectional TEM lamella using a Thermo Scientific Helios G4 PFIB UXe Dual Beam equipped with an EasyLift manipulator. Standard lift-out technique was used to prepare the TEM lamella. [100]-oriented TEM specimen was prepared using Ar-ion milling (Fischione Model 1010). A 4 keV ion beam with the angle of incidence set at 5° and followed by a 1 keV ion beam with the angle of incidence at 2° was used to thin down the specimen. All the TEM lamellae were heated to 130 °C in vacuum for 8 hours to remove organic contaminants from the surface before being inserted into the microscope column.

Scanning transmission electron microscope (STEM) imaging was performed using an aberration-corrected Nion UltraSTEM 100 operated at 100 kV with a convergence semi-angle of 30 mrad. HAADF-STEM images were acquired using an annular dark-field detector with inner and outer collection semi-angles of 80 and 200 mrad, respectively.

To interpret the intensity variation in a STEM image, multi-slice simulations were carried out on the structures of $Sr_{9/8}TiS_3$ and hypothetical stoichiometric $SrTiS_3$. The structure of $Sr_{9/8}TiS_3$ was obtained from the structural refinement and the structure of $SrTiS_3$ was obtained after optimization using DFT. STEM-HAADF simulations were performed using the multi-slice method as implemented in μSTEM[55]. Thermal scattering was included in our simulations through the phonon-excitation model proposed by Forbes et al.[56] The sample thickness was set to 15 nm and the defocus value was set to 10 Å to obtain good agreement in intensity profiles with the experimental data. We performed the simulations using an aberration-free probe with an accelerating voltage of 100 kV and a convergence semi-angle of 30 mrad. The inner and outer collection angles for the HAADF detector were set to 80 and 200 mrad, respectively.

**Theoretical calculations**

We performed density-functional theory (DFT) calculations using projected augmented-wave potentials[57] as implemented in the Vienna Ab initio Simulation Package (VASP)[58,59]. The Perdew-Burke-Ernzerhof (PBE) functional within the generalized gradient approximation (GGA)[60] was used to describe the exchange-correlation interactions. A plane-wave basis with an energy cutoff of 600 eV and an energy convergence criterion of $10^{-8}$ eV for the electronic convergence were applied. A *k*-point spacing of 0.025 $Å^{-1}$ was chosen for both the structure optimization and total-energy calculation. The criterion for structural optimization was set such that all forces on the atoms were less than $10^{-4}$ eV/Å. To increase the localization of Ti-3*d* electrons, we used DFT + *U* calculations[51]. An effective on-site Hubbard $U = 3.0$ eV was used for the Ti-3*d* electrons. Furthermore, we considered magnetic configurations for both stoichiometric $SrTiS_3$ ($P2_1$) and modulated $Sr_{9/8}TiS_3$ ($R3c$) lattices. The special quasi-random structure (SQS) model implemented in the alloy theoretic automatic toolkit[61,62] was used to generate best approximations of randomness in the paramagnetic configuration. The visualization of band decomposed charge density was performed for valence electrons within 0.3 eV below Fermi energy. After running convergence tests, we set the total number of energy bands (NBANDS = 736 for primitive cell) to be 2.5 times as many as the number of valence bands for the dielectric function calculations which covers the energy transition up to 62 eV.

## Acknowledgements

The work at UW-Madison was supported by ONR, with award no. N00014-20-1-2297. The work at USC and WUStL were supported, in part, by an ARO MURI program with award no. W911NF-21-1-0327, and the National Science Foundation (NSF) of the United States under grant numbers DMR-2122070 and DMR-2122071. J.R. acknowledges support from the Army Research Office under Award No. W911NF-19-1-0137, and an Air Force Office of Scientific Research grant no. FA9550-22-1-0117. This research used resources of the Advanced Light Source, which is a DOE Office of Science User Facility under contract no. DE-AC02-05CH11231. J.R., B.Z., and H.C. gratefully acknowledge the use of Core Center for Excellence in Nano Imaging (CNI), University of Southern California for some of the sample preparation and characterization studies. R.M. acknowledges NSF for partial support through grant DMR-2145797. M.K. and H.M. acknowledge the use of facilities and instrumentation at the UW-Madison Wisconsin Centers for Nanoscale Technology (WCNT) partially supported by the NSF through the University of Wisconsin Materials Research Science and Engineering Center (DMR-1720415). STEM characterization was performed at the Center for Nanophase Materials Sciences and X-ray structural work at the Spallation Neutron Source, both of which are US Department of Energy, Office of Science User Facility operated by Oak Ridge National Laboratory. This work used computational resources through allocation DMR160007 from the Advanced Cyberinfrastructure Coordination Ecosystem: Services & Support (ACCESS) program, which is supported by NSF grants #2138259, #2138286, #2138307, #2137603, and #2138296.

## Author contributions

M.A.K., R.M. and J.R. conceived and supervised the research. H.M., J.S. and M.A.K. identified the large optical anisotropy. G.R., G.Y.J. and R.M. performed the structural modulation studies. H.M. and J.S. performed IR spectroscopy. H.M., J.S. and C.W. performed the ellipsometry studies. B.Z., S.S., and S.N. grew the crystals and performed structural and chemical characterization. B.Z., N.S., S.J.T., and B.C.C. performed single-crystal XRD measurements. G.Y.J., G.R., J.C. and R.M. performed the theoretical calculations. G.R. and A.S.T. performed the STEM experiments and analysis with assistance from J.A.H., M.C., H.C., and R.M. All authors discussed the results. H.M., M.A.K., G.R., B.Z, J.R., and R.M. wrote the manuscript with contributions from all co-authors.

## Competing interests

The authors declare no conflict of interest.

## Data availability

Data presented in the main text and supplementary information are open access and can be found on Zenodo[63].

# Supplementary Information

# Colossal optical anisotropy from atomic-scale modulations

Hongyan Mei[1†], Guodong Ren[2†], Boyang Zhao[3†], Jad Salman[1†], Gwan Yeong Jung[4], Huandong Chen[3], Shantanu Singh[3], Arashdeep S. Thind[2], John Cavin[5], Jordan A. Hachtel[6], Miaofang Chi[6], Shanyuan Niu[3], Graham Joe[1], Chenghao Wan[1,7], Nick Settineri[8], Simon J. Teat[8], Bryan C. Chakoumakos[9], Jayakanth Ravichandran[3,10,11*], Rohan Mishra[2,4*], Mikhail A. Kats[1,7*]

[1] *Department of Electrical and Computer Engineering, University of Wisconsin-Madison, Madison, WI 53706, USA*
[2] *Institute of Materials Science and Engineering, Washington University in St. Louis, St. Louis, MO 63130, USA*
[3] *Mork Family Department of Chemical Engineering and Materials Science, University of Southern California, Los Angeles, CA 90089, USA*
[4] *Department of Mechanical Engineering and Material Science, Washington University in St. Louis, St. Louis, MO 63130, USA*
[5] *Department of Physics, Washington University in St. Louis, St. Louis, MO 63130, USA*
[6] *Center for Nanophase Materials Sciences, Oak Ridge National Laboratory, Oak Ridge, TN 37831, USA*
[7] *Department of Materials Science and Engineering, University of Wisconsin-Madison, Madison, WI 53706, USA*
[8] *Advanced Light Source, Lawrence Berkeley National Laboratory, Berkeley, CA 94720, USA*
[9] *Neutron Scattering Division, Oak Ridge National Laboratory, Oak Ridge TN 37831, USA*
[10] *Ming Hsieh Department of Electrical Engineering, University of Southern California, Los Angeles, CA 90089, USA*
[11] *Core Center for Excellence in NanoImaging, University of Southern California, Los Angeles, CA 90089, USA*
[†] *These authors contributed equally to this work*
**Email: mkats@wisc.edu, rmishra@wustl.edu, jayakanr@usc.edu*

## Contents

**Section 1. Single crystal diffraction: data collection, data reduction and structure refinements**

We conducted single crystal diffraction (SC-XRD) at atomic resolution to extract the structure of $Sr_{9/8}TiS_3$. Needle-shaped $Sr_{9/8}TiS_3$ crystals (~0.1×0.1×0.2 mm$^3$, reported in Table S1, S2 and S3) and platelet-shaped $Sr_{9/8}TiS_3$ crystals (~0.3×0.1×0.5 mm$^3$) were measured. Needle-shaped crystal pieces are well resolved as single domain [0*kl*, *h*0*l*, *hk*0 precession maps in Fig. S1(a-c)] $R\bar{3}c$ ($R_1$=0.0358) or merohedral twined $R3c$ ($R_1$=0.0322) and $R32$ ($R_1$=0.0324), compared in Table S1. Platelet crystals show non-merohedral twins [0*kl*, *h*0*l*, *hk*0 precession maps in Fig. S1(d-f)] along the *c*-axis, represented by their highly twined reflections (expanded view Fig. S1(d) inset) in the *h*0*l* and 0*kl* precession maps but clean reflections in *hk*0 [Fig. S1(d, e)]. Regardless of the non-merohedral twins, platelet crystals show consistent $R3c$ space group when treated as pseudo-merohedral twins ($R_1$ increases to 0.0667).

Based on the characterized structure modulation [Fig. 2(a, b)] of $Sr_{9/8}TiS_3$, we converted the 3D reciprocal space into commensurate (3+1)D $hkl_1l_2$ superstructure of $TiS_3$ ($hkl_10$) and Sr ($hk0l_2$) sublattices, as labeled in Fig. S1(a-c). $TiS_3$ sublattice shows systematic extinctions for $-h+k+l_1=3n$, which matches with $R\bar{3}$ sublattice space group, while Sr sublattice with $h-k=3n$, $l_2=2n$ for $00l_2$ and $h0l_2$ systematic extinctions reveals a $P\bar{3}c1$ sublattice space group. A superspace group best matched both sublattices is then ascribed as $R\bar{3}$m(00*g*)0s.

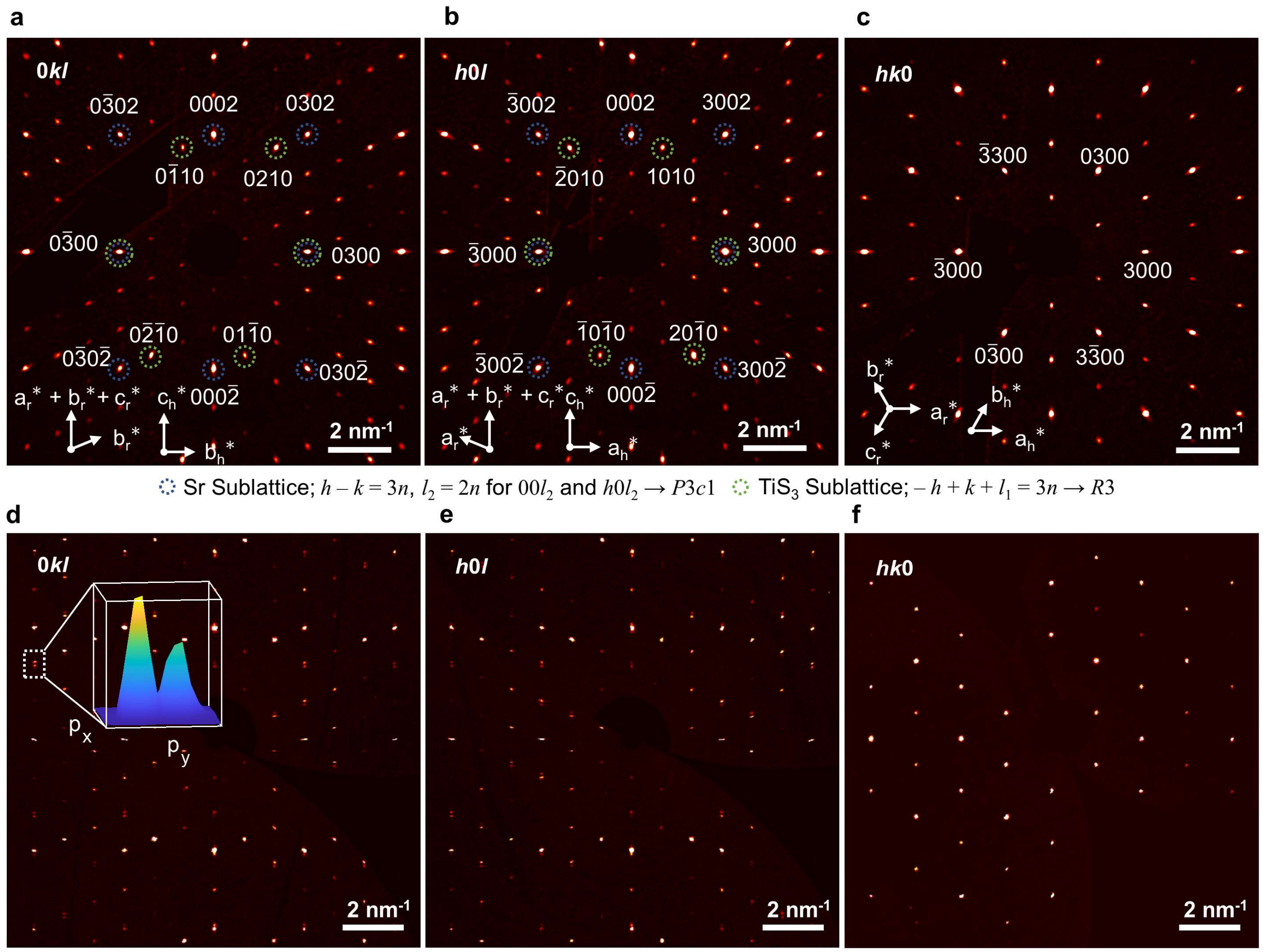

Non-merohedral Twined $Sr_{9/8}TiS_3$ Crystal

**Supplementary Figure S1. Single crystal X-ray diffraction of $Sr_{9/8}TiS_3$. a-c.** 0*kl*, *h*0*l* and *hk*0 planes of the reciprocal precession map of needle-shaped $Sr_{9/8}TiS_3$. Rhombohedral principal reciprocal axes ($a_r^*$, $b_r^*$, $c_r^*$), equivalent hexagonal unit cell reciprocal axes ($a_h^*$, $b_h^*$, $c_h^*$) of *R*3*c* space group are labeled. Main reflections are indexed as (3+1)D $hkl_1l_2$ for sublattice space group determination. **d-e.** 0*kl*, *h*0*l* and *hk*0 planes of the reciprocal precession map of platelet-shaped $Sr_{9/8}TiS_3$ in a single ω scan. Certain reflections in 0*kl* and *h*0*l* show non-merohedral twin as the inset illustrates.

**Supplementary Table S1.** Data collection, intensity statistics, and refinement statistics of single crystal diffraction on $Sr_{1.125}TiS_3$.

| | $Sr_{1.125}TiS_3$ | | | | | | | | |
|---|---|---|---|---|---|---|---|---|---|
| **Temperature** | 260 K | | | | | | | | |
| **Radiation Type** | Mo *K*α | | | | | | | | |
| **Wavelength** | 0.71073 Å | | | | | | | | |
| **Space group** | *R*3*c* | | | *R*-3*c* | | | *R*32 | | |
| **Cell dimensions** | | | | | | | | | |
| *a*, *b*, *c* (Å) | 11.4555(4), 11.4555(4), 47.6287(18) | | | | | | | | |
| α, β, γ (°) | 90, 90, 120 | | | | | | | | |
| Volume (Å$^3$) | 5412.9(4) | | | | | | | | |
| Density (g/cm$^3$) | 3.573 | | | | | | | | |
| **Reduction & Scaling** | Point group symmetry: -3*m*1 | | | | | | | | |
| Resolution (Å) | inf-1.47 | 1.47-1.15 | 1.15-1.00 | 1.00-0.90 | 0.90-0.84 | 0.84-0.78 | 0.78-0.74 | 0.74-0.71 | 0.71-0.68 |
| Obs. Reflections | 3999 | 4474 | 3526 | 3003 | 2575 | 2270 | 2122 | 1970 | 1328 |
| Completeness (%) | 99.6 | 100 | 100 | 100 | 100 | 100 | 100 | 100 | 97.7 |
| Redundancy | 15.8 | 17.7 | 13.9 | 11.9 | 10.2 | 9 | 8.4 | 7.8 | 5.2 |
| Mean $I$ / σ$I$ | 61.92 | 52.15 | 40.07 | 31.95 | 20.86 | 18.33 | 16.87 | 13.12 | 9.84 |
| $R_{int}$ | 0.031 | 0.035 | 0.046 | 0.056 | 0.102 | 0.113 | 0.109 | 0.147 | 0.137 |
| $R_σ$ | 0.012 | 0.013 | 0.017 | 0.022 | 0.035 | 0.042 | 0.045 | 0.064 | 0.088 |
| **Reflections** | $I > 2σI$ | | | | | | | | |
| $R_σ$, $R_{int}$ | 0.0343, 0.0501 | | | 0.0233, 0.0512 | | | 0.0340, 0.0501 | | |
| $θ_{min}$, $θ_{full}$, $θ_{max}$ (°) | 2.097, 25.242, 30.504 | | | 2.097, 25.242, 30.504 | | | 2.097, 25.242, 30.504 | | |
| Friedel fraction full | 0.938 | | | ~ | | | 1.000 | | |
| **Refinement** | | | | | | | | | |
| Weighting | $w = 1/[σ^2(F_o^2) + (0.0217P)^2 + 48.6113P]$, where $P = (F_o^2 + 2F_c^2) / 3$ | | | | | | | | |
| Resolution (Å) | 50.0 - 0.70 | | | 50.0 - 0.70 | | | 50.0 - 0.70 | | |
| No. of reflections | 3465 | | | 1846 | | | 3690 | | |
| No. of $I > 2σI$ | 2872 | | | 1622 | | | 2823 | | |
| No. parameters | 126 | | | 65 | | | 128 | | |
| No. constraints | 1 | | | 0 | | | 0 | | |

| | | | |
|---|---|---|---|
| $R_1$, $wR_2$ [all data] | 0.0468, 0.0659 | 0.0439, 0.0677 | 0.0532, 0.0679 |
| $R_1$, $wR_2$ [$I > 4\sigma I$] | 0.0322, 0.0593 | 0.0358, 0.0637 | 0.0324, 0.0591 |
| *GoF* | 1.070 | 1.095 | 1.073 |
| Twin Matrix | $\begin{pmatrix} 1 & 0 & 0 \\ 0 & 1 & 0 \\ 0 & 0 & 1 \end{pmatrix}, \begin{pmatrix} 0 & 1 & 0 \\ 1 & 0 & 0 \\ 0 & 0 & -1 \end{pmatrix}$ | $\begin{pmatrix} 1 & 0 & 0 \\ 0 & 1 & 0 \\ 0 & 0 & 1 \end{pmatrix}$ | $\begin{pmatrix} 1 & 0 & 0 \\ 0 & 1 & 0 \\ 0 & 0 & 1 \end{pmatrix}, \begin{pmatrix} -1 & 0 & 0 \\ 0 & -1 & 0 \\ 0 & 0 & -1 \end{pmatrix}$ |
| BASF | 0.50697, 0.49303 | 1 | 0.54625, 0.45375 |

**Supplementary Table S2.** Atomic coordinates and equivalent isotropic atomic displacement parameters for *R*3*c* $Sr_{1.125}TiS_3$.

| Atom | *x/a* | *y/b* | *z/c* | $U_{iso}$ (Å$^2$) | Occupancy | Symmetry Order |
|---|---|---|---|---|---|---|
| Sr01 | 0.6627(2) | 0.69470(19) | 0.47341(3) | 0.0155(5) | 1 | 1 |
| Sr02 | 0.33830(19) | 0.30412(18) | 0.42312(3) | 0.0110(4) | 1 | 1 |
| Sr03 | -0.01735(15) | 0.31715(15) | 0.36505(8) | 0.01654(14) | 1 | 1 |
| Ti04 | 0.333333 | 0.666667 | 0.67383(14) | 0.0101(11) | 1 | 3 |
| Ti05 | 0.333333 | 0.666667 | 0.73435(14) | 0.0119(12) | 1 | 3 |
| Ti06 | 0.333333 | 0.666667 | 0.42931(14) | 0.0095(11) | 1 | 3 |
| Ti07 | 0.333333 | 0.666667 | 0.61528(19) | 0.0073(3) | 1 | 3 |
| Ti08 | 0.333333 | 0.666667 | 0.3646(2) | 0.0130(3) | 1 | 3 |
| Ti09 | 0.333333 | 0.666667 | 0.30027(14) | 0.0095(11) | 1 | 3 |
| Ti0A | 0.333333 | 0.666667 | 0.49553(13) | 0.0078(11) | 1 | 3 |
| Ti0B | 0.333333 | 0.666667 | 0.55666(13) | 0.0058(10) | 1 | 3 |
| S00C | 0.1646(5) | 0.6675(5) | 0.70429(10) | 0.0092(9) | 1 | 1 |
| S00D | 0.3253(5) | 0.4973(5) | 0.26680(10) | 0.0090(9) | 1 | 1 |
| S00E | 0.3094(5) | 0.8278(5) | 0.64432(10) | 0.0078(8) | 1 | 1 |
| S00F | 0.5036(5) | 0.6683(5) | 0.52486(10) | 0.0107(9) | 1 | 1 |
| S00G | 0.1478(5) | 0.6437(5) | 0.58459(10) | 0.0102(9) | 1 | 1 |
| S00H | 0.3403(5) | 0.5053(5) | 0.46237(10) | 0.0118(10) | 1 | 1 |
| S00I | 0.2345(6) | 0.4788(5) | 0.33233(12) | 0.0255(12) | 1 | 1 |
| S00J | 0.4295(6) | 0.5745(5) | 0.39707(12) | 0.0179(10) | 1 | 1 |

**Supplementary Table S3.** Anisotropic atomic displacement parameters for *R*3*c* $Sr_{1.125}TiS_3$.

| Atom | $U_{11}$ (Å$^2$) | $U_{22}$ (Å$^2$) | $U_{33}$ (Å$^2$) | $U_{23}$ (Å$^2$) | $U_{13}$ (Å$^2$) | $U_{12}$ (Å$^2$) |
|---|---|---|---|---|---|---|
| Sr01 | 0.0135(9) | 0.0116(9) | 0.0176(11) | -0.0026(7) | 0.0078(6) | 0.0033(7) |
| Sr02 | 0.0133(8) | 0.0113(8) | 0.0119(9) | 0.0036(6) | 0.0042(6) | 0.0087(7) |
| Sr03 | 0.0114(9) | 0.0142(10) | 0.0192(3) | 0.0069(9) | -0.0017(9) | 0.0028(2) |
| Ti04 | 0.0099(15) | 0.0099(15) | 0.011(3) | 0 | 0 | 0.0050(7) |
| Ti05 | 0.0111(17) | 0.0111(17) | 0.013(3) | 0 | 0 | 0.0055(8) |
| Ti06 | 0.0088(16) | 0.0088(16) | 0.011(3) | 0 | 0 | 0.0044(8) |
| Ti07 | 0.0071(4) | 0.0071(4) | 0.0077(8) | 0 | 0 | 0.0035(2) |
| Ti08 | 0.0134(5) | 0.0134(5) | 0.0122(8) | 0 | 0 | 0.0067(2) |
| Ti09 | 0.0087(16) | 0.0087(16) | 0.011(3) | 0 | 0 | 0.0044(8) |
| Ti0A | 0.0094(16) | 0.0094(16) | 0.005(3) | 0 | 0 | 0.0047(8) |
| Ti0B | 0.0078(15) | 0.0078(15) | 0.002(2) | 0 | 0 | 0.0039(7) |
| S00C | 0.008(2) | 0.015(2) | 0.010(2) | -0.0017(16) | 0.0022(16) | 0.0096(17) |
| S00D | 0.013(2) | 0.005(2) | 0.008(2) | 0.0013(15) | 0.0032(16) | 0.0032(16) |
| S00E | 0.0107(18) | 0.007(2) | 0.007(2) | 0.0010(14) | -0.0021(16) | 0.0049(16) |
| S00F | 0.010(2) | 0.016(2) | 0.006(2) | 0.0006(16) | -0.0018(16) | 0.0059(17) |
| S00G | 0.010(2) | 0.018(2) | 0.005(2) | -0.0018(17) | -0.0027(16) | 0.0091(18) |
| S00H | 0.017(2) | 0.009(2) | 0.010(2) | -0.0042(16) | -0.0035(17) | 0.0069(18) |
| S00I | 0.039(3) | 0.006(2) | 0.013(2) | -0.0026(17) | 0.013(2) | -0.0026(19) |
| S00J | 0.030(2) | 0.018(2) | 0.015(2) | 0.0085(17) | 0.0102(19) | 0.0185(18) |

## Section 2. Extraction of the anisotropic complex refractive index

We combined ellipsometry (UV-NIR) with polarized spectroscopy (NIR-MIR) to extract the complex refractive index of $Sr_{9/8}TiS_3$. The $S_{9/8}TiS_3$ crystal was suspended over air with no substrate and the ellipsometry measurements were made using a VASE ellipsometer (J.A. Woollam Co.) outfitted with focusing probes. The VASE ellipsometer covers the spectral range from 210 nm to 2500 nm and the focusing probes can provide a 200 μm beam diameter at normal incidence, with the beam diameter increasing as a function of $1/\cos(\theta_i)$ in the horizontal direction ($\theta_i$ is the angle of incidence). Three ellipsometry measurements were made with the crystal *c*-axis parallel to the plane of incidence, perpendicular to the plane of incidence, and at an off-axis angle (~30°). The three measurements were all made at 55° from the surface normal.

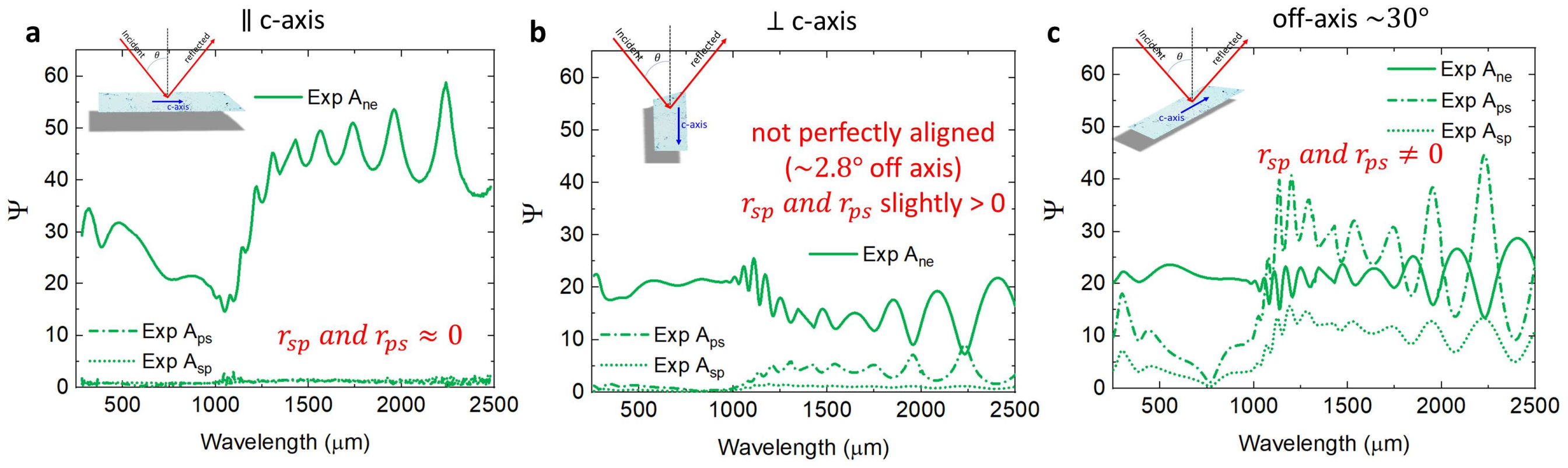

**Supplementary Figure S2.** Raw ellipsometry data for $Sr_{9/8}TiS_3$ at $\theta_i = 55°$ with the crystal *c*-axis parallel to the plane of incidence **a**, perpendicular to the plane of incidence **b**, and at an off-axis angle **c**. Shown here are only the Ψ components of the measured ellipsometric data, which are related to the amplitude of the ratio of the Fresnel coefficients. Δ values are not shown.

Fig. S2 shows the raw measured ellipsometry data ($A_{ps}, A_{sp}, A_{ne}$) for all three measurements for $Sr_{9/8}TiS_3$. $A_{ne}$ is the ratio between the two complex Fresnel coefficients, which can be described by:

$$A_{ne} = \rho = \frac{r_{pp}}{r_{ss}} = \tan(\Psi_{A_{ne}})\, e^{i\Delta_{Ane}}$$

where Ψ and Δ are the two quantities measured by the ellipsometer. For anisotropic materials, $A_{ps}$ and $A_{sp}$ are used to capture cross-polarization interactions, which are defined as:

$$A_{ps} = \frac{r_{ps}}{r_{pp}} = \tan\left(\Psi_{A_{ps}}\right) e^{i\Delta_{A_{ps}}}$$

$$A_{sp} = \frac{r_{sp}}{r_{ss}} = \tan\left(\Psi_{A_{sp}}\right) e^{i\Delta_{A_{sp}}}$$

All three datasets were fit simultaneously using a single optical model with uniaxial birefringence. Here, optical properties perpendicular and parallel to the *c*-axis contain independent oscillators. By incorporating the polarization-resolved reflectance and transmittance of $Sr_{9/8}TiS_3$ to the ellipsometry data, the optical model could be extended up to the detection limit of the FTIR (17 μm). Thus, seven independent optical measurements (three ellipsometry measurements, two reflectance measurements, and two transmittance measurements) could all be fit simultaneously to a single anisotropic optical model to yield a full set of refractive indices for the samples spanning the 210 nm through 17 μm. Fig. S4 shows the final fit oscillator models along the ordinary and extraordinary directions and Fig. S5 shows the good agreement between the experimental data and fitted data. The total oscillator model for each direction is the sum of individual Kramers-Kronig-consistent oscillators. Table S4 lists the oscillators used in the anisotropic optical model for both the ordinary and extraordinary directions.

The fitted thickness for $Sr_{9/8}TiS_3$ is 3.2 μm, which is in good agreement with the SEM cross section measurement in Fig. S3b.

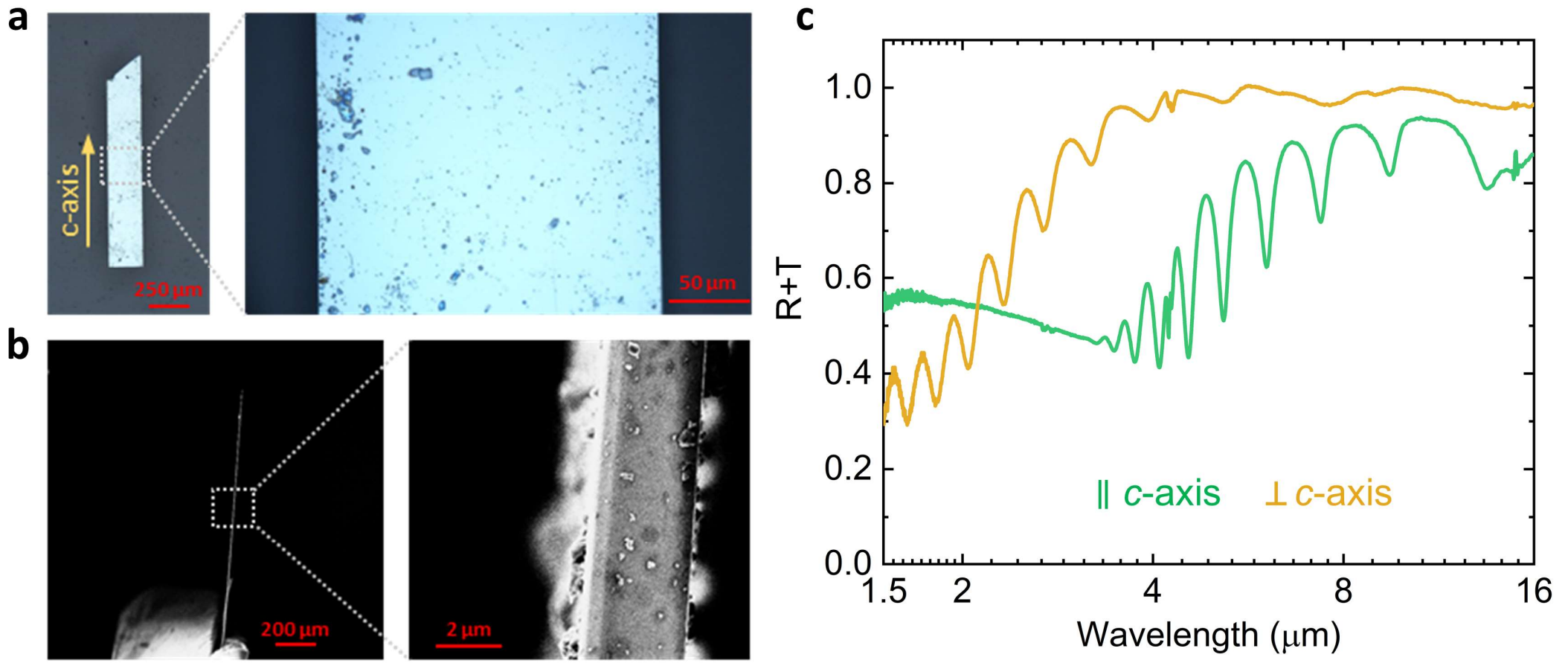

**Supplementary Figure S3. a.** optical image of a $Sr_{9/8}TiS_3$ crystal plate. **b.** SEM cross-section image of a $Sr_{9/8}TiS_3$ crystal plate, showing the thickness of ~3.2 micron. **c.** sum of the measured reflectance and transmittance, showing the low loss region of $Sr_{9/8}TiS_3$.

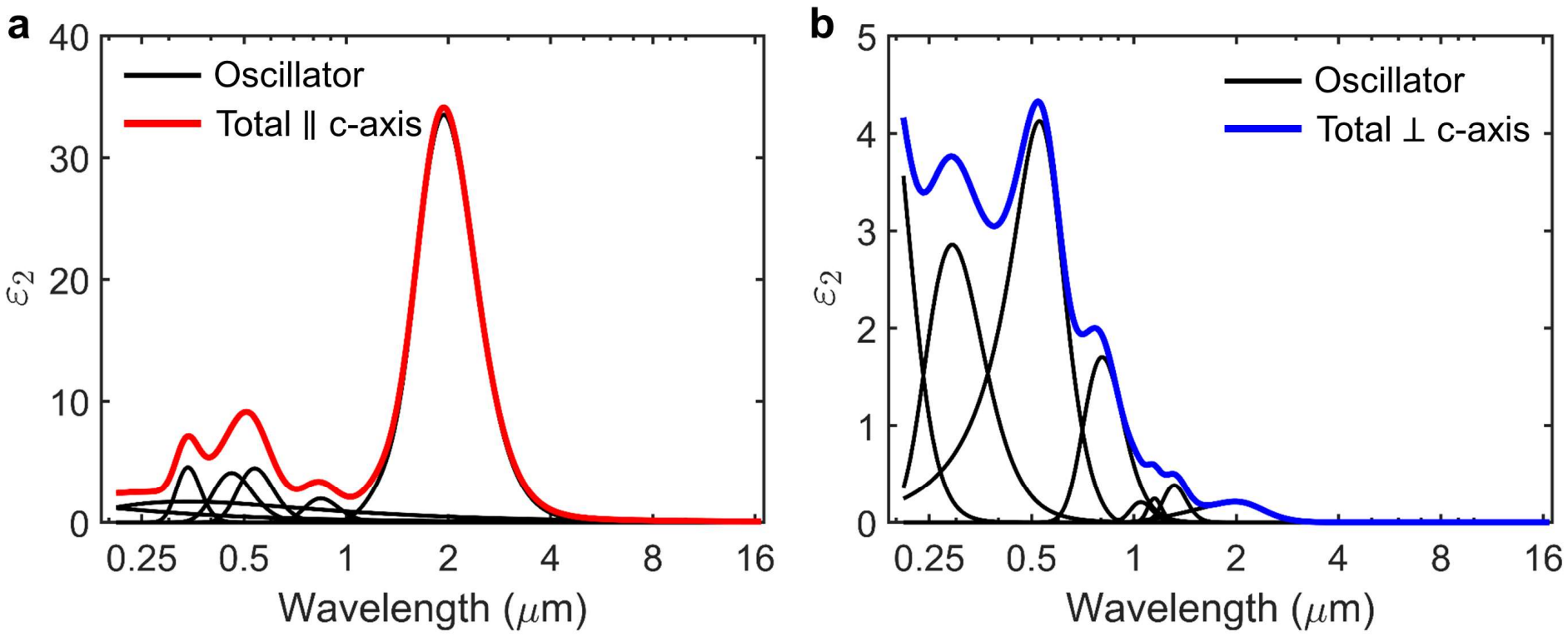

**Supplementary Figure S4.** The $\varepsilon_2$ values calculated from the final optical oscillator model used to fit to the ellipsometry data of $Sr_{9/8}TiS_3$. The optical oscillator model is built on Kramers-Kronig consistent oscillators.

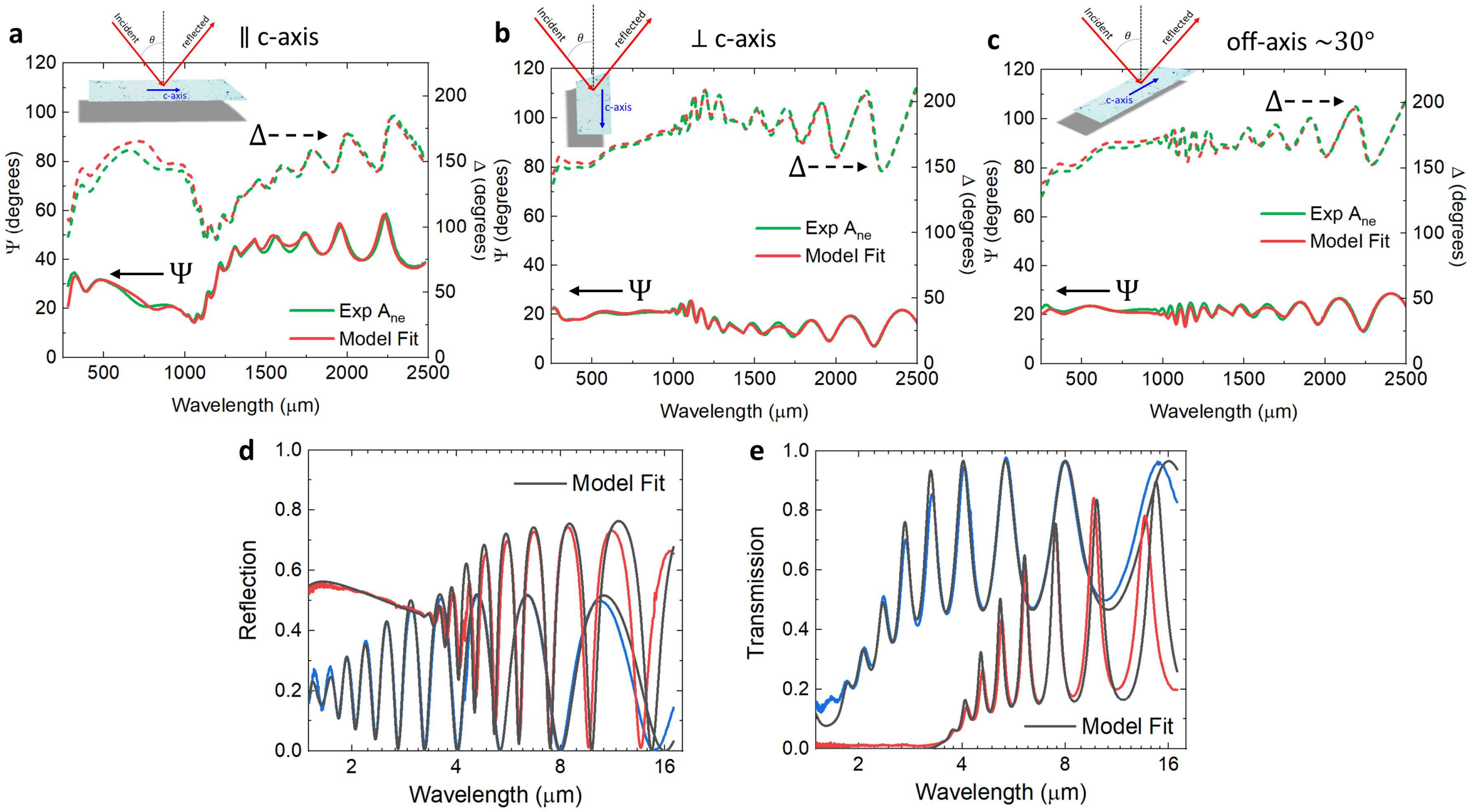

**Supplementary Figure S5. a-c.** Final fitted model data of $Sr_{9/8}TiS_3$, in red, compared to measured ellipsometry data (Ψ and Δ), in green. Final anisotropic optical model values are described in Table S4. **d-e.** Final fitted model data of $Sr_{9/8}TiS_3$, in black, compared to measured polarization-resolved FTIR reflectance and transmittance data, in red and blue.

The extracted refractive indices for $Sr_{9/8}TiS_3$ are shown in Fig. 1(d) in the main text and colossal birefringence is observed spanning the entire mid- through far-infrared ranges within a broad low-loss window.

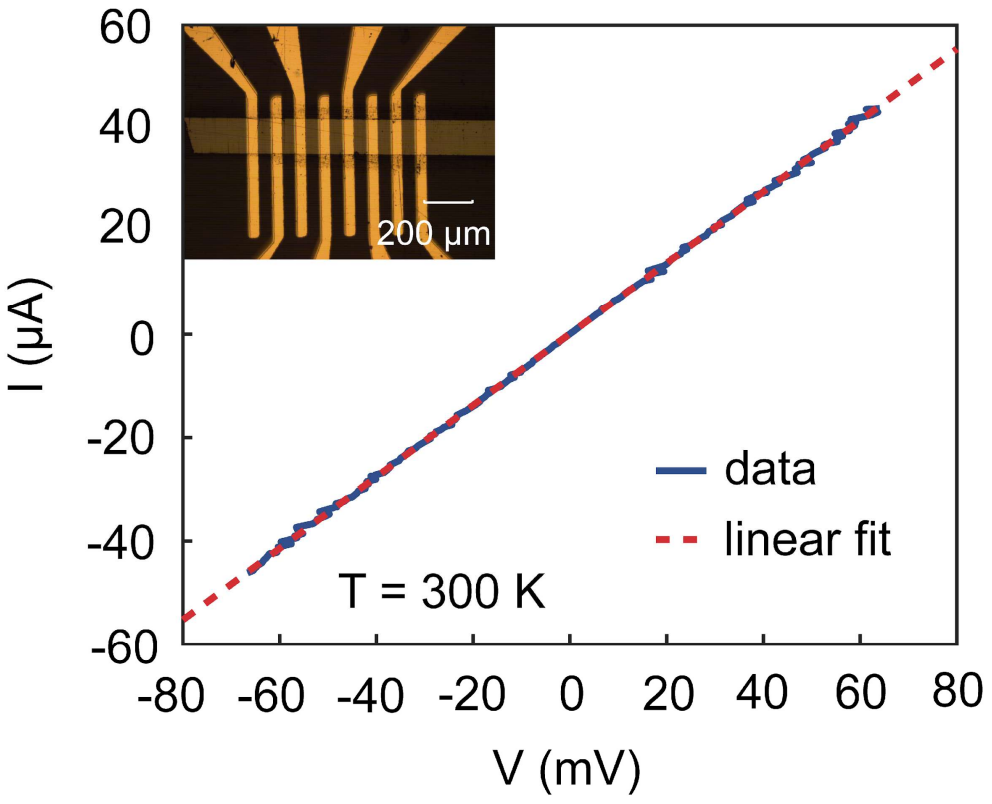

**Supplementary Figure S6.** Four-probe current-voltage (I-V) characteristics of a $Sr_{9/8}TiS_3$ crystal, with the inset showing optical microscopic image of the $Sr_{9/8}TiS_3$ device.

We carried out DC current-voltage characterization on the $Sr_{9/8}TiS_3$ crystal with four-probe geometry at room temperature to extract the electrical conductivity as illustrated in Fig. S6, which is close to ~2.7 $\Omega \cdot cm$. From the previous measurements [S1], we know that $BaTiS_3$ has a contact resistivity of ~0.4 $\Omega \cdot cm$ and a carrier concentration of ~1.1×10$^{18}$ cm$^{-3}$. Therefore, we anticipate that the $Sr_{9/8}TiS_3$ crystal has a carrier concentration of ~1.6×10$^{17}$ cm$^{-3}$ (we assume the mobility of $BaTiS_3$ and $Sr_{9/8}TiS_3$ are similar). From the DFT calculation, we can get the effective mass of $Sr_{9/8}TiS_3$, which is 1.5. Thus, according to the Drude model [S2], we can calculate the plasmon wavelength of our $Sr_{9/8}TiS_3$ sample, which is ~100 μm.

$$\omega_p^2 = \frac{Nq^2}{\varepsilon_0 m^*}, \lambda_p = \frac{2\pi c}{\omega_p}$$

**Supplementary Table S4.** Oscillators and the fitted parameters of the $Sr_{9/8}TiS_3$ crystal in Fig. S3**.**

| Oscillator | Parameters (eV) | | |
|---|---|---|---|
| Gaussian<br>$\varepsilon_{Gaus} = \varepsilon_1 + i\varepsilon_2$<br>$\varepsilon_2 = A_n e^{\left(\frac{E-E_n}{\sigma}\right)^2} - A_n e^{-\left(\frac{E+E_n}{\sigma}\right)^2}$<br>$\varepsilon_1 = \frac{2}{\pi} P \int_0^\infty \frac{\xi \varepsilon_2(\xi)}{\xi^2 - E^2} d\xi$<br>Where, $\sigma = \frac{Br_n}{2\sqrt{\ln(2)}}$, $P$ is the Cauchy Principal Value | Ordinary<br>$A_1$ = 1.2389<br>$A_2$ = 2.8591<br>$A_3$ = 4.5825<br>$A_4$ = 0.25625<br>$A_5$ = 0.21926<br>$A_6$ = 0.4238<br>Extraordinary<br>$A_7$ = 4.1011<br>$A_8$ = 4.4872<br>$A_9$ = 1.709<br>$A_{10}$ = 2.8654<br>$A_{11}$ = 4.5852<br>$A_{12}$ = 1.404 | Ordinary<br>$E_1$ = 1.537<br>$E_2$ = 4.2407<br>$E_3$ = 6.5818<br>$E_4$ = 0.75646<br>$E_5$ = 0.94097<br>$E_6$ = 0.00065<br>Extraordinary<br>$E_7$ = 2.6967<br>$E_8$ = 2.3077<br>$E_9$ = 1.3753<br>$E_{10}$ = 1.8896<br>$E_{11}$ = 3.6265<br>$E_{12}$ = 8.3814 | Ordinary<br>$Br_1$ = 0.48721<br>$Br_2$ = 1.9164<br>$Br_3$ = 2.2495<br>$Br_4$ = 0.12458<br>$Br_5$ = 0.2041<br>$Br_6$ = 0.17878<br>Extraordinary<br>$Br_7$ = 0.82102<br>$Br_8$ = 0.65113<br>$Br_9$ = 0.26131<br>$Br_{10}$ = 8.1843<br>$Br_{11}$ = 0.69007<br>$Br_{12}$ = 9.8278 |
| Tauc-Lorentz<br>$\varepsilon_{TL} = \varepsilon_1 + i\varepsilon_2$<br>$\varepsilon_2 = \frac{A_n Eo_n C_n (E - Eg_n)^2}{(E^2 - Eo_n^2)^2 + C_n^2 E^2} \cdot \frac{1}{E}, for\ E > Eg_n$<br>$\varepsilon_2 = 0, for\ E \leq Eg_n$<br>$\varepsilon_1 = \frac{2}{\pi} P \int_{Eg_n}^\infty \frac{\xi \varepsilon_2(\xi)}{\xi^2 - E^2} d\xi$ | Ordinary<br>$A_1$ = 26.696, $Eo_1$ = 2.2384, $C_1$ = 1.142, $Eg_1$ = 1.3129 | | |
| Parametric Semiconductor<br>See WVASE manual by J.A. Woollam Co. [S3] | Ordinary (Psemi-M0)<br>$A_1$ = 0.31885, $Eo_1$ = 0.51766, $B_1$ = 0.088998, $WR_1$ = 0.86135, $PR_1$ = 0.62087, $AR_1$ = 0.35457, $O2R_1$ = 1<br>Extraordinary (Psemi-Tri)<br>$A_1$ = 135.04, $Ec_1$ = 0.5936, $B_1$ = 0.1323, $WL_1$ = 0.10456, $WR_1$ = 0.55047, $AL_1$ = 0.0021511, $AR_1$ = 0.043094 | | |

The reproducibility of the ellipsometry fitting is tested by measuring another $Sr_{9/8}TiS_3$ crystal with a different thickness, which was synthesized in 2020. As shown in Fig. S7(a), the thickness is ~19.4 μm. We carried out the similar measurements (ellipsometry and the polarization-resolved FTIR reflectance) and followed the similar fitting procedures. The extracted complex refractive-index values of two $Sr_{9/8}TiS_3$ crystals are displayed in Fig. S7(b), showing good consistency between two separate fittings. We note that the measurements and fits were performed by two different authors at different times, increasing our confidence in the measurements.

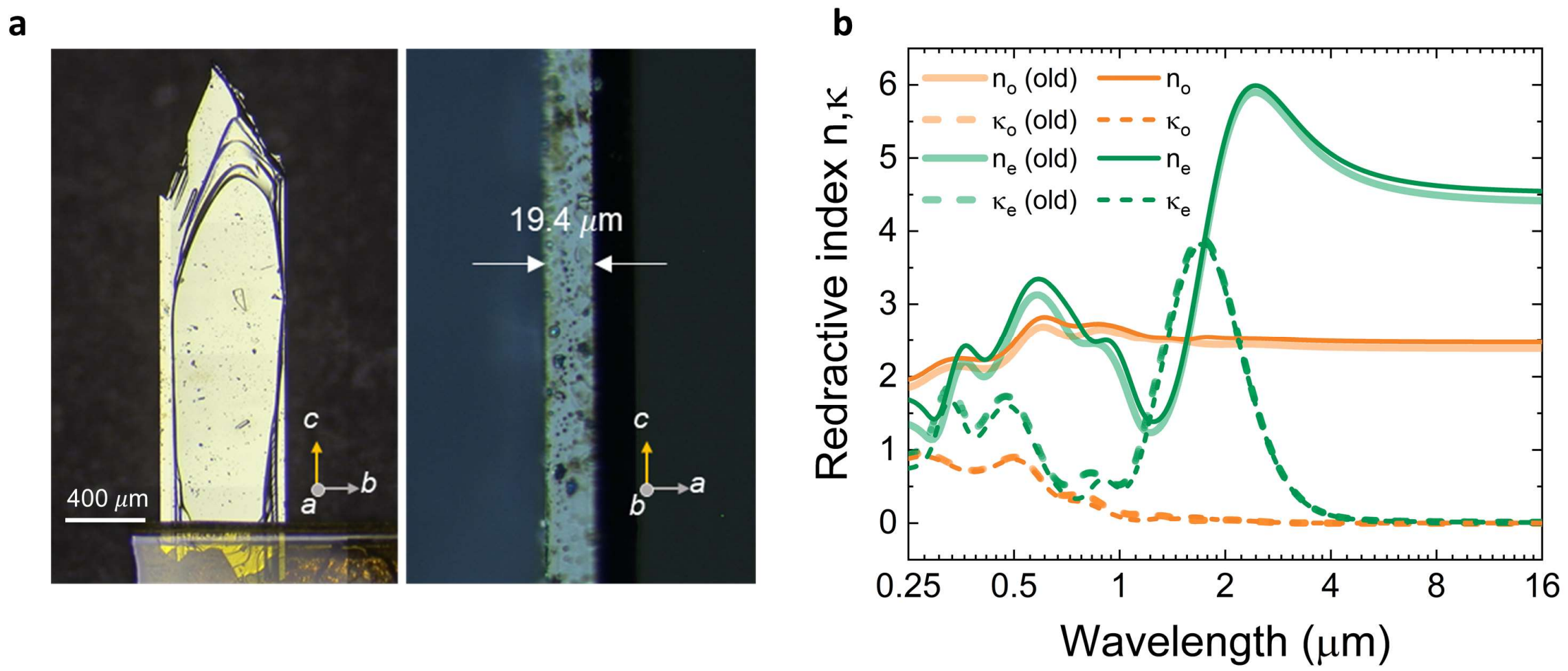

**Supplementary Figure S7. a.** Optical images of the $Sr_{9/8}TiS_3$ crystal with a thickness of ~19.4 μm. The lateral dimensions are approximately 500 by 1200 μm (scale bar shown). **b.** Extracted complex refractive-index values of two different $Sr_{9/8}TiS_3$ crystals. Dark green and orange lines show the results of the current 3.2-μm-thick $Sr_{9/8}TiS_3$ crystal in the main paper, and the light green and orange lines show the results of the old 19.4-μm-thick $Sr_{9/8}TiS_3$ crystal in **a.** Two sets of values appear to be well matched.

**Supplementary Table S5.** Oscillators and the fitted parameters of the $Sr_{9/8}TiS_3$ crystal in Fig. S7**.**

| **Oscillator** | **Parameters (eV)** | | |
|---|---|---|---|
| Gaussian<br>$\varepsilon_{Gaus} = \varepsilon_1 + i\varepsilon_2$<br>$\varepsilon_2 = A_n e^{\left(\frac{E-E_n}{\sigma}\right)^2} - A_n e^{-\left(\frac{E+E_n}{\sigma}\right)^2}$<br>$\varepsilon_1 = \frac{2}{\pi} P \int_0^\infty \frac{\xi \varepsilon_2(\xi)}{\xi^2 - E^2} d\xi$<br>Where, $\sigma = \frac{Br_n}{2\sqrt{\ln(2)}}$, $P$ is the Cauchy Principal Value | Ordinary | Ordinary | Ordinary |
| | $A_1 = 1.7046$ | $E = 1.537$ | $Br_1 = 0.48197$ |
| | $A_2 = 2.8591$ | $E_2 = 4.2407$ | $Br_2 = 1.9164$ |
| | $A_3 = 4.5825$ | $E_3 = 6.5818$ | $Br_3 = 2.2495$ |
| | $A_4 = 0.20854$ | $E_4 = 1.1826$ | $Br_4 = 0.20466$ |
| | $A_5 = 0.247$ | $E_5 = 1.0783$ | $Br_5 = 0.11757$ |
| | $A_6 = 0.38514$ | $E_6 = 0.94299$ | $Br_6 = 0.17152$ |
| | Extraordinary | Extraordinary | Extraordinary |
| | $A_7 = 1.404$ | E7 = 8.3814 | $Br_7 = 9.8278$ |
| | $A_8 = 4.5852$ | E8 = 3.6265 | $Br_8 = 0.69007$ |
| | $A_9 = 4.1011$ | E9 = 2.6967 | $Br_9 = 0.84102$ |
| | $A_{10} = 4.4872$ | E10 = 2.3077 | $Br_{10} = 0.65113$ |
| | $A_{11} = 2.8654$ | E11 = 1.8896 | $Br_{11} = 8.1843$ |
| | $A_{12} = 1.9774$ | E12 = 1.4741 | $Br_{12} = 0.40093$ |
| Tauc-Lorentz<br>$\varepsilon_{TL} = \varepsilon_1 + i\varepsilon_2$ | Ordinary<br>$A_1 = 25.654$, $Eo_1 = 2.2384$, $C_1 = 1.1106$, $Eg_1 = 1.3129$ | | |

| | |
|---|---|
| $$\varepsilon_2 = \frac{A_n Eo_n C_n (E - Eg_n)^2}{(E^2 - Eo_n^2)^2 + C_n^2 E^2} \cdot \frac{1}{E}, \ for\ E > Eg_n$$ $$\varepsilon_2 = 0,\ for\ E \le Eg_n$$ $$\varepsilon_1 = \frac{2}{\pi} P \int_{Eg_n}^{\infty} \frac{\xi \varepsilon_2(\xi)}{\xi^2 - E^2} d\xi$$ | |
| Parametric Semiconductor | Ordinary (Psemi-M0)<br>$A_1$ = 0.32204, $Eo_1$ = 0.52297, $B_1$ = 0.084786, $WR_1$ = 0.86135, $PR_1$ = 0.62087, $AR_1$ = 0.35457, $O2R_1$ = 1<br>Extraordinary (Psemi-Tri)<br>$A_1$ = 135.04, $Ec_1$ = 0.5936, $B_1$ = 0.1323, $WL_1$ = 0.10456, $WR_1$ = 0.55047, $AL_1$ = 0.0021511, $AR_1$ = 0.043094 |

## Section 3. Stoichiometric $SrTiS_3$

The highest symmetry structure of stoichiometric $SrTiS_3$ in Materials Project database [S4] has a space group of $P6_3/mmc$, which is computed to be metallic at both the PBE and PBE+$U$ levels ($U$ = 3 eV for Ti atoms). The calculated phonon dispersion of this structure shows significant soft phonon modes at both the zone center and zone boundaries (see Figure S8a), implying the presence of dynamic instabilities. We froze the soft phonon modes at the Γ point (including $\Gamma_2^-$, $\Gamma_2^+$, $\Gamma_5^-$, and $\Gamma_5^+$ modes) to arrive at the $P2_1$ phase, as shown in Fig. S8b. $P2_1$-$SrTiS_3$ has a finite bad gap of 0.84 eV at the PBE+$U$ level, and has a lower energy than $P6_3/mmc$-$SrTiS_3$ by 44.3 meV/atom. We then performed electronic structure calculations and STEM image simulations on the hypothetical $P2_1$-$SrTiS_3$, for comparison with non-stoichiometric $Sr_{9/8}TiS_3$.

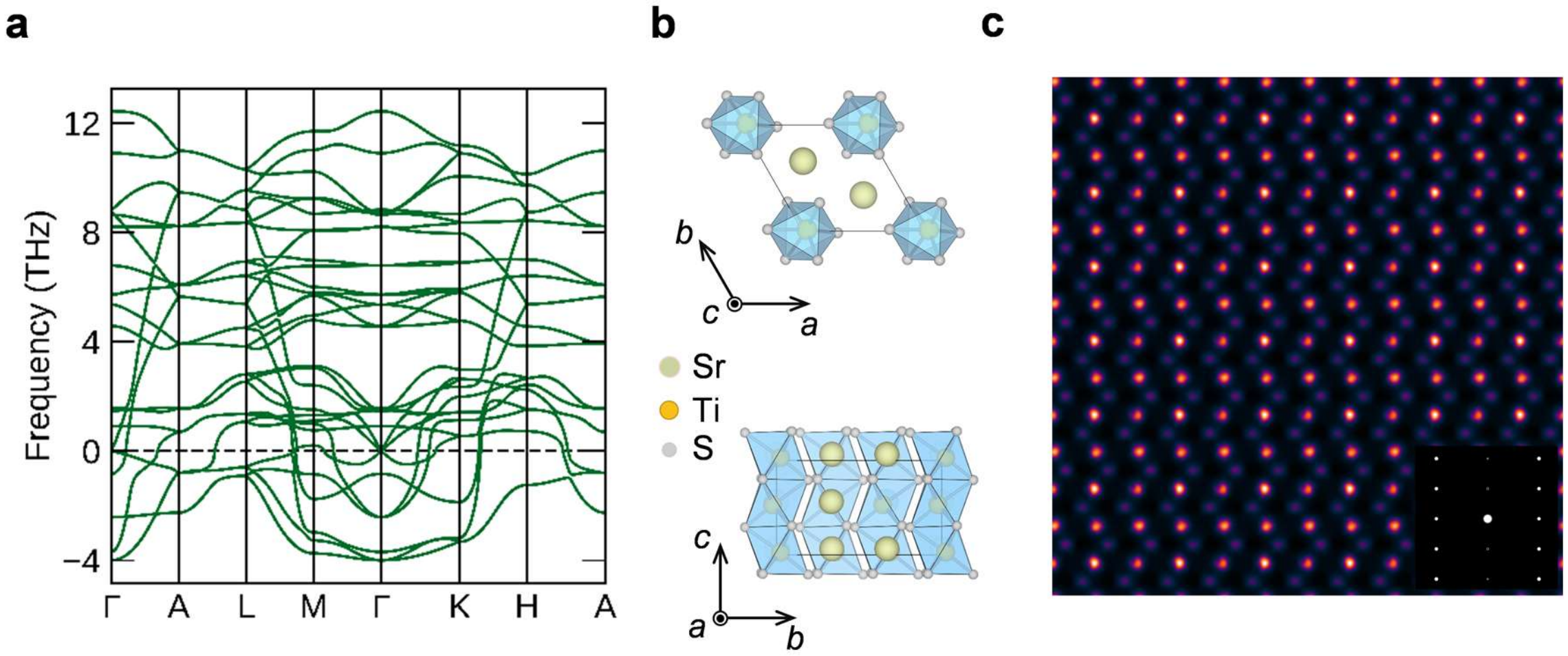

**Supplementary Figure S8. Structure of stoichiometric $SrTiS_3$. a**. Phonon dispersion of stochiometric $P6_3/mmc$-$SrTiS_3$ showing lattice instability due to the presence of soft-phonon modes. **b**. Schematic representing lower-symmetry $P2_1$-$SrTiS_3$ viewed from [001] and [100] orientations, where the $TiS_6$ units form face-sharing octahedral chains running along the *c*-axis. **d.** Simulated HAADF-STEM image of $P2_1$-$SrTiS_3$. The inset shows the simulated electron diffraction pattern along the [100] orientation.

**Section 4. Thermodynamic stability through convex-hull analysis**

We evaluated the thermodynamic stability of both stoichiometric $SrTiS_3$ ($P2_1$) and modulated $Sr_{9/8}TiS_3$ ($R3c$) by constructing the convex hull with respect to the formation energy of their possible decomposition products. The convex hull connects phases that have a formation energy lower than any other phase or linear combination of phases at the respective compositions. If a phase has a formation energy above the hull, $E_{hull}$, it is considered metastable, as the system can lower its energy (by $E_{hull}$) by decomposing into products that are on the hull [S5]. For the Sr-Ti-S system, we calculated the total energy ($E_{total}$) of all the thermodynamically stable phases ($E_{hull} = 0$) available in the Materials Project [S4]. With all phases listed in the Table S6 being considered, we constructed the convex hull through the grand canonical linear programming (GCLP) method [S6]. The GCLP minimizes the free energy of a mixture at a given composition in the Sr-Ti-S phase space to identify the combination of thermodynamically equilibrium phases:

$$\Delta G = \sum_i f_i \Delta E_f^i,$$

where, $\Delta G$ is the free energy of the compound with the desired composition, $f_i$ is the molar fraction of competing phases, $\Delta E_f^i$ is the formation energy of competing phases. We calculated $\Delta E_f^i$ of each phase using:

$$\Delta E_f^i = E_{\mathrm{Sr}_x\mathrm{Ti}_y\mathrm{S}_z} - xE_{\mathrm{Sr}} - yE_{\mathrm{Ti}} - zE_{\mathrm{S}},$$

where $E_{\mathrm{Sr}_x\mathrm{Ti}_y\mathrm{S}_z}$, $E_{\mathrm{Sr}}$, $E_{\mathrm{Ti}}$ and $E_{\mathrm{S}}$ are the total energy of each phase from DFT calculations; while $x$, $y$ and $z$ are atomic fraction of each element in the compounds.

For $SrTiS_3$ and $Sr_{9/8}TiS_3$, we find the decomposition products that minimize the total energy are:

$$\mathrm{SrTiS_3} \rightarrow \mathrm{SrS} + \mathrm{TiS_2}$$
$$\mathrm{Sr_9Ti_8S_{24}} \rightarrow 9\mathrm{SrS} + 4.5\mathrm{TiS_2} + 0.5\mathrm{Ti_7S_{12}}$$

As shown in the tabulated dataset Table S6 for the Sr-Ti-S system, modulated $Sr_{9/8}TiS_3$ is thermodynamically stable with formation energy on the hull ($E_{hull} = 0$), while stoichiometric $SrTiS_3$ is metastable having formation energy above hull ($E_{hull} = 44.975$ meV/atom).

**Supplementary Table S6.** Database for convex hull construction

| Formula | Space group | $E_{total}$/eV | Atoms/unit | $\Delta E_f$ (eV/atom) | $E_{hull}$ (meV/atom) |
|---|---|---|---|---|---|
| Sr | *R*-3*m* | -5.017 | 3 | 0 | |
| Ti | *P*6/*mmm* | -16.578 | 3 | 0 | |
| S | *P*2/*c* | -132.103 | 32 | 0 | |
| SrS | *Fm*-3*m* | -10.113 | 2 | -2.156 | |
| $SrS_3$ | *Aba*2 | -36.283 | 8 | -1.021 | |
| $Ti_2S$ | *Pnnm* | -218.902 | 36 | -1.020 | |
| $Ti_2S_3$ | *C*2/*m* | -61.002 | 10 | -1.413 | |
| $Ti_5S_8$ | *C*2/*m* | -79.046 | 13 | -1.415 | |
| $Ti_7S_{12}$ | *P*-1 | -229.359 | 38 | -1.393 | |
| TiS | *P*-6*m*2 | -12.506 | 2 | -1.426 | |
| $TiS_2$ | *P*-3*m*1 | -17.849 | 3 | -1.355 | |
| $TiS_3$ | $P2_1/m$ | -43.744 | 8 | -0.990 | |
| $SrTiS_3$ | $P2_1$ | -55.474 | 10 | -1.631 | 44.975 |
| $Sr_{9/8}TiS_3$ | *R*3c | -457.439 | 82 | -1.717 | 0 |

**Section 5. Crystal structure of modulated $Sr_{9/8}TiS_3$**

To accommodate the excess Sr atoms in $Sr_{9/8}TiS_3$ ($R3c$), certain $TiS_6$ units undergo a periodic trigonal twist distortion resulting in the formation of $TiS_6$ trigonal-prismatic polyhedra from the octahedral polyhedra in $SrTiS_3$. Fig. S9(b, c) show two building blocks of the modulated $Sr_{9/8}TiS_3$ lattice. Compared to face-shared octahedral $TiS_6$ units (O) represented in blue, the pseudo-trigonal prismatic units (T) in orange have elongated Ti–S bonds and have a ~30 ˚ rotation of $S_3$ triangles around the *c*-axis. As shown in Fig. S9(a, b), Sr atoms show displacive modulation about the average position along both *c*-axis and *ab* plane, which leads to the twist distortion of $TiS_6$ units accompanied with altered Ti–Ti distances along the *c*-axis. When projected along the [100] zone axis, Ti atoms periodically overlap with Sr as highlighted with dashed ellipses in Figure S9a.

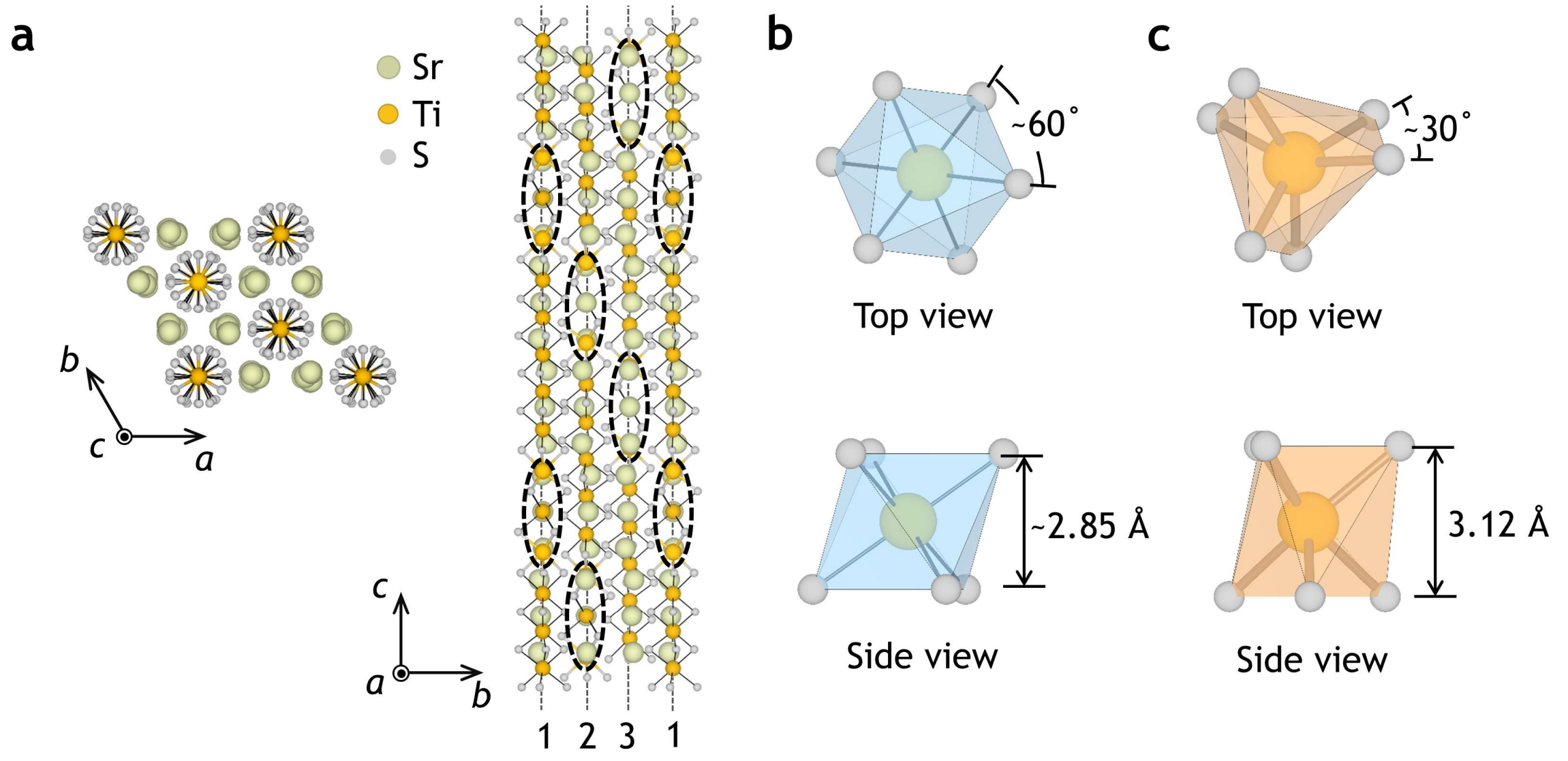

**Supplementary Figure S9. Structural modulations in $Sr_{9/8}TiS_3$. a.** Schematic representing modulated $Sr_{9/8}TiS_3$. Three identical $TiS_6$ chains (1, 2, 3) are shown. They have a commensurate stacking sequence of $[-(T-O-T)-(O)_5-]_2$, which defines the modulation periodicity. The dashed ellipses indicate the periodic (T−O−T) segments, where Ti atoms overlap with Sr atoms, when projected along [100] zone axis. **b, c.** Schematic representing building blocks for modulated $Sr_{9/8}TiS_3$ with top and side views. Face-sharing octahedra (O) are highlighted in blue, and pseudo-trigonal prismatic (T) units are represented in orange.

The displacive modulation of Sr resulting from structural modulations has been confirmed by measuring horizontal offsets of neighboring Sr atoms along *c*-axis in the HAADF-STEM images viewed along the [100] zone axis, as shown in Fig. S10. The position of Sr atomic columns were determined by applying a 2D Gaussian peak fitting alogrithm [S7]. As shown in Fig. S10, the Sr configuration in the experimental HAADF image shows good match with that in the simulated image obtained using the refined $Sr_{9/8}TiS_3$ lattice, which verifies the modulation periodicity of $Sr_{9/8}TiS_3$ as solved from X-ray diffraction.

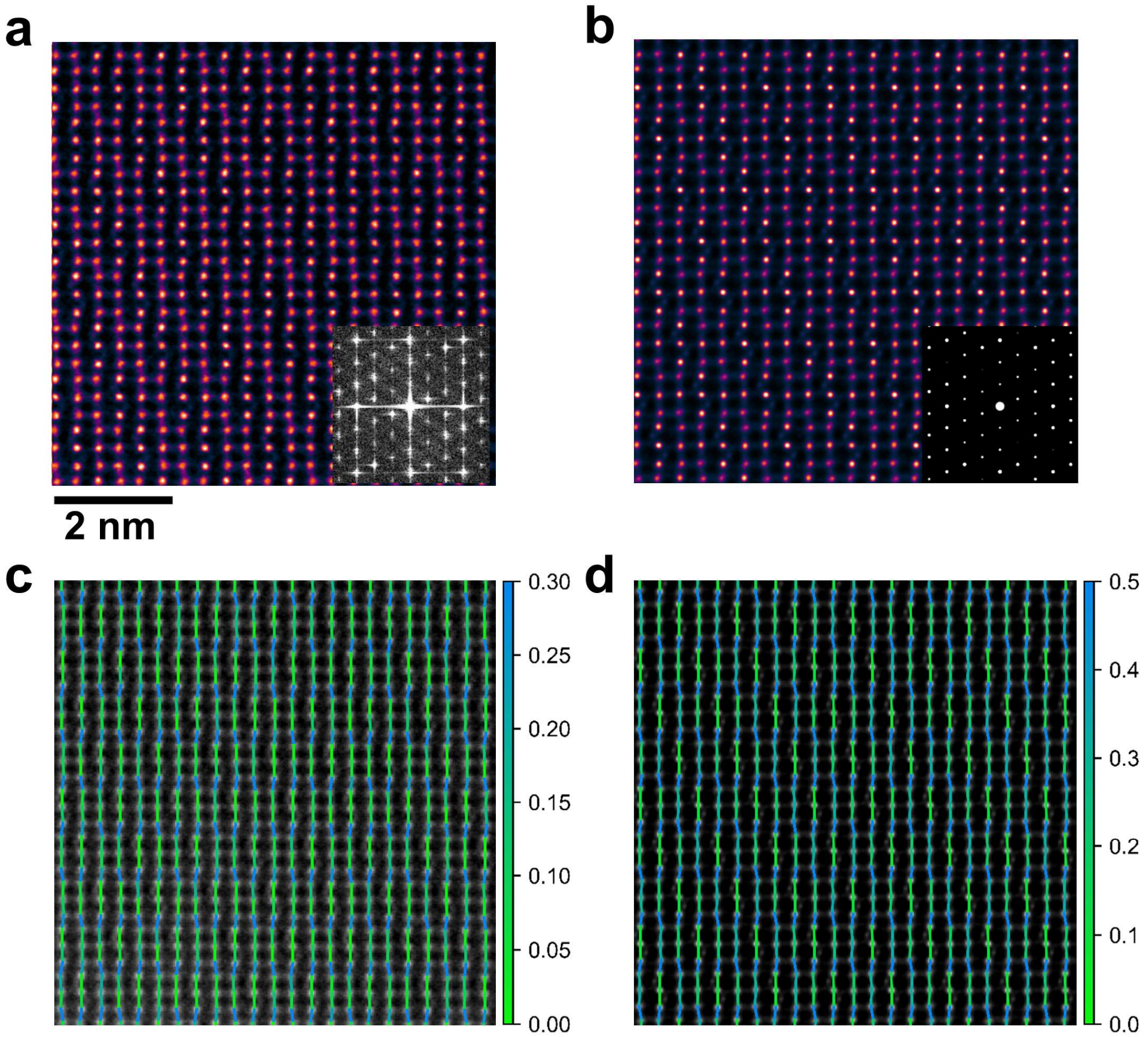

**Supplementary Figure S10. Experimental and simulated HAADF-STEM images of modulated $Sr_{9/8}TiS_3$. a.** Direct experimental observation of modulations in $Sr_{9/8}TiS_3$ viewed along [100] zone axis from a large field of view HAADF image. The inset is the corresponding fast Fourier transform (FFT) pattern. **b.** Simulated HAADF-STEM image of modulated $Sr_{9/8}TiS_3$. The inset shows the simulated electron diffraction pattern. **c, d.** Maps of structural modulation periodicity along the *c*-axis by measuring horizontal offsets of the neighboring atomic columns from **a,b**. The color of connecting lines represents the magnitude of deviation along the horizontal direction between neighboring atoms. The length scale of the color bar is in the units of Å.

**Section 6. Crystal-field splitting with modulations in $Sr_{9/8}TiS_3$**

The hybridization between S-*p* and Ti-*d* states is shown in Figure S11 using a color scale that quantifies the projected wavefunction character in the band structure. Sr states were excluded from the visualization as they are nominally empty in the energy range shown in Figure S11, and are present at higher energies in the conduction band. We are also showing the projected density of states (PDOS). $SrTiS_3$ exhibits a *p*-*d* transition from the S *p*-valence band to empty Ti *d*-states. Meanwhile, $Sr_{9/8}TiS_3$ shows a *d*-*d* transition due to the occupied Ti $d_{z^2}$-states within the octahedral blocks.

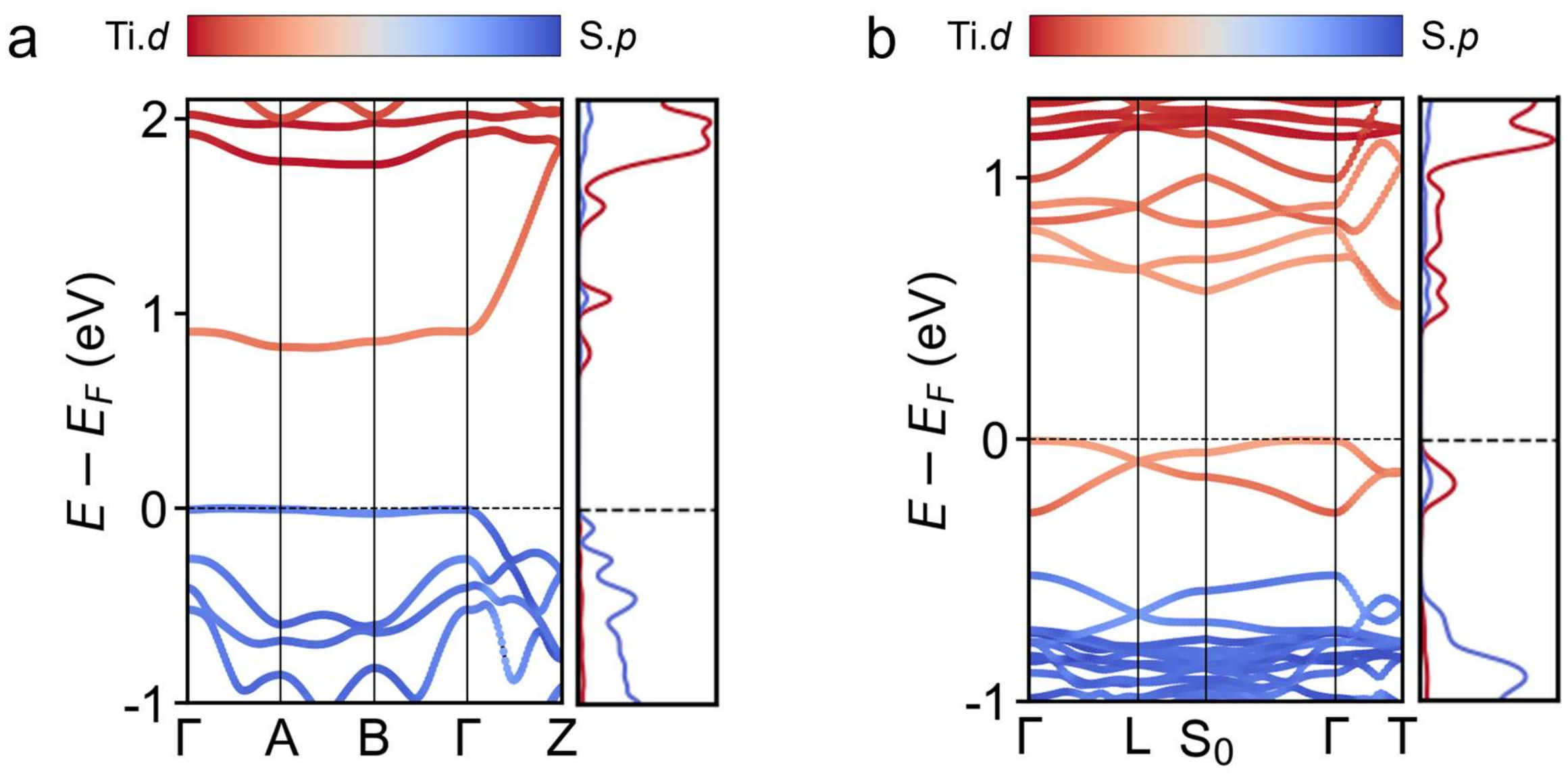

**Supplementary Figure S11.** (Left panel) Band structure with projected wave-function character and (right panel) element- and orbital-projected density of states (PDOS) showing the hybridization of S-*p* and Ti-*d* states in **a.** $SrTiS_3$ and in **b.** $Sr_{9/8}TiS_3$.

The energy levels of *d* orbitals in Ti cations are affected by the crystal-field splitting arising from the surrounding negatively charged anions [S8]. As schematically illustrated in Fig. S12, in a face-sharing octahedra geometry, the Ti cation has point symmetry of $D3_d$ with Ti–S bonds pointing diagonally in octahedral coordination. The degeneracy of five-fold 3*d* orbitals of Ti cations is removed by considering the orientation of each orbital with respect to the charge distribution on neighboring $S^{2-}$ anions. The nominally empty $d^0$ orbitals of $Ti^{4+}$ cation split into single $a_{1g}$ orbital with dominant $d_{z^2}$ character along the *c*-axis, doubly degenerate $e_g^{\pi}$ orbitals in *ab* plane, and another set of doubly degenerate $e_g^{\sigma}$ orbitals pointing along the Ti-S bonds [S9]. The empty Ti *3d* states primarily form the conduction bands. The $a_{1g}$ state has the least overlap with the S-*3p* states, and being an antibonding state, it is lowest in energy. The $e_g^{\sigma}$ states have the most overlap, and have the highest energy. The $e_g^{\pi}$ states are at intermediate energy.

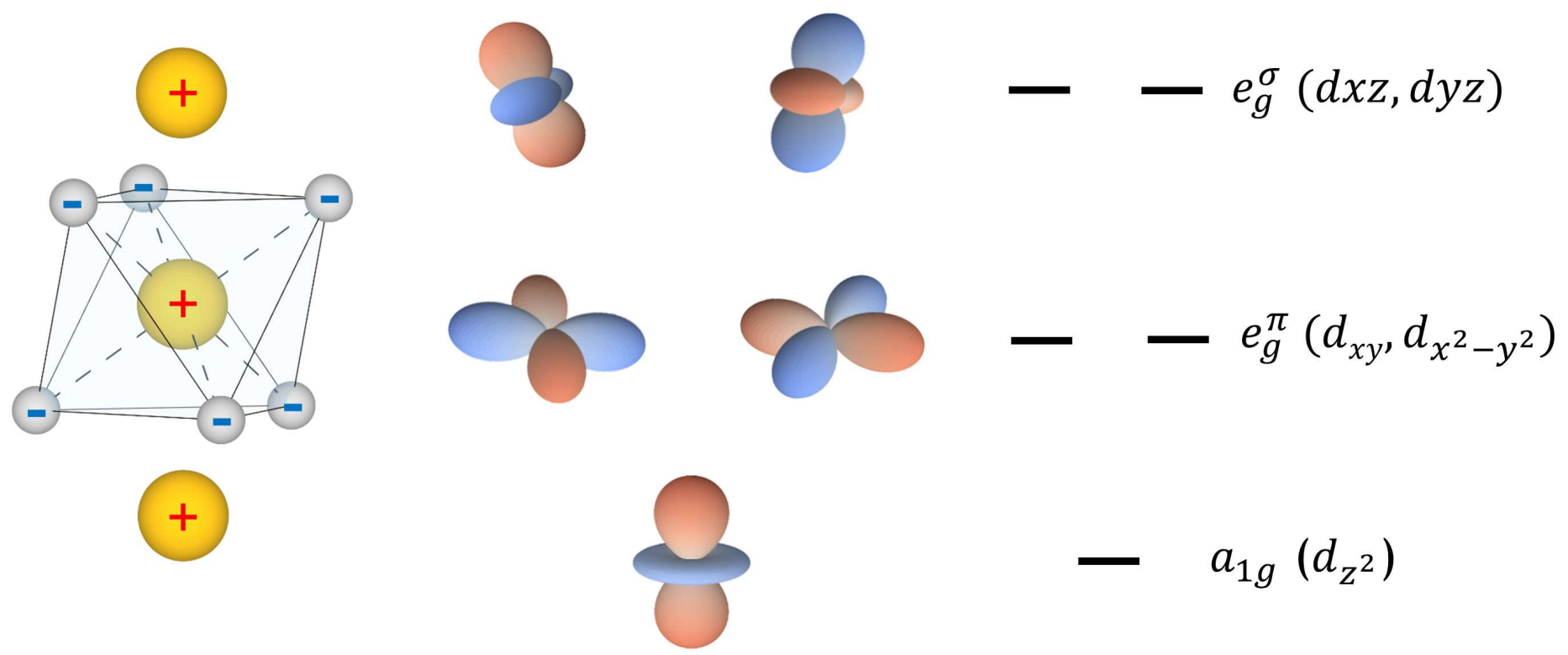

**Supplementary Figure S12. Face-sharing octahedra geometry and crystal-field splitting.** The $TiS_6$ octahedron shows $D_{3d}$ symmetry with a three-fold rotational axis along $TiS_6$ chain direction. The Ti–S bonds are represented by dashed lines. The crystal field of face-sharing octahedra splits the 3*d* orbitals into an $a_{1g}$ singlet state, and $e_g^{\pi}$ and $e_g^{\sigma}$ doublets. The coordinate for orbitals is defined as the *z*-axis is along $TiS_6$ chain direction.

While ideal face-sharing octahedra with $D_{3d}$ symmetry splits the Ti-3*d* orbitals into two doublets ($e_g^{\pi}$ and $e_g^{\sigma}$) and a singlet ($a_{1g}$), the symmetry reduction by trigonal distortion and the connectivity of neighboring octahedra further alters the splitting of the *d* orbitals by lifting their degeneracies [S9]. The $Sr_{9/8}TiS_3$ crystal shows structural modulation with a stacking sequence of building blocks of T and O along the *c*-axis. Therefore, one can expect different orbital structures at different $TiS_6$ segments in the $Sr_{9/8}TiS_3$ lattice.

DFT calculations were performed on the modulated $Sr_{9/8}TiS_3$ lattice with the structure obtained from SC-XRD. The fully optimized lattice has been compared with the structure as resolved from SC-XRD in terms of space group symmetry and structural modulation periodicity. In Fig. S13a, both XRD refinement and DFT calculations show Ti–Ti distance modulation in $TiS_6$ chains with longer distance within the (T−O−T) segments, and shorter distance within the $(O)_5$ segments.

Here, we explore how the changes in $TiS_6$ configuration alters the crystal-field splitting of the 3*d* orbitals of Ti. In Fig. S13b, we show the projected density of states (PDOS) of 3*d* states of Ti atoms in octahedra (O) and pseudo-trigonal prismatic (T) sites. The center of mass of PDOS is used to estimate the energy levels of the different orbitals. The $d_{z^2}$ ($a_{1g}$) states on the O sites have lower energy than that of the T sites, and are preferentially occupied by the additional electrons introduced by excess Sr. Meanwhile, the remaining *d* states have almost the same energy levels on both the T and O sites. Therefore, the trigonal twist distortion mainly causes the increase of energy level in $d_{z^2}$ ($a_{1g}$) derived states.

As suggested by XRD refinement and DFT calculations, the $Sr_{9/8}TiS_3$ lattice has structural modulation accompanied by Ti–Ti distance variation along the *c*-axis. Upon visualizing the isosurfaces of the band-decomposed charge density (see electronic structure in Fig. 3 in the main text), we find selective occupation of $d_{z^2}$ states on O sites that have shorter Ti–Ti distance, with the Ti atoms at the middle of the $(O)_5$ segments showing the highest occupancy of $d_{z^2}$ states as they have the shortest Ti–Ti distance, as shown in Fig. S13c.

In conclusion, the trigonal twist distortion together with Ti–Ti distance modulation further splits the level of $d_{z^2}$ ($a_{1g}$) derived states. As schematically shown in Fig. S13d, unlike the ideal face-sharing octahedral crystal-field with $D_{3d}$ symmetry, structural modulations in the $Sr_{9/8}TiS_3$ lattice lower the $d_{z^2}$ ($a_{1g}$) states of the Ti atoms within the $(O)_5$ segments, which are occupied by the electron from the excess Sr. This reforms the band gap $\Delta E_g$, which is now the energy difference between the occupied $a_{1g}$ states and the empty $a_{1g}{}'$ states.

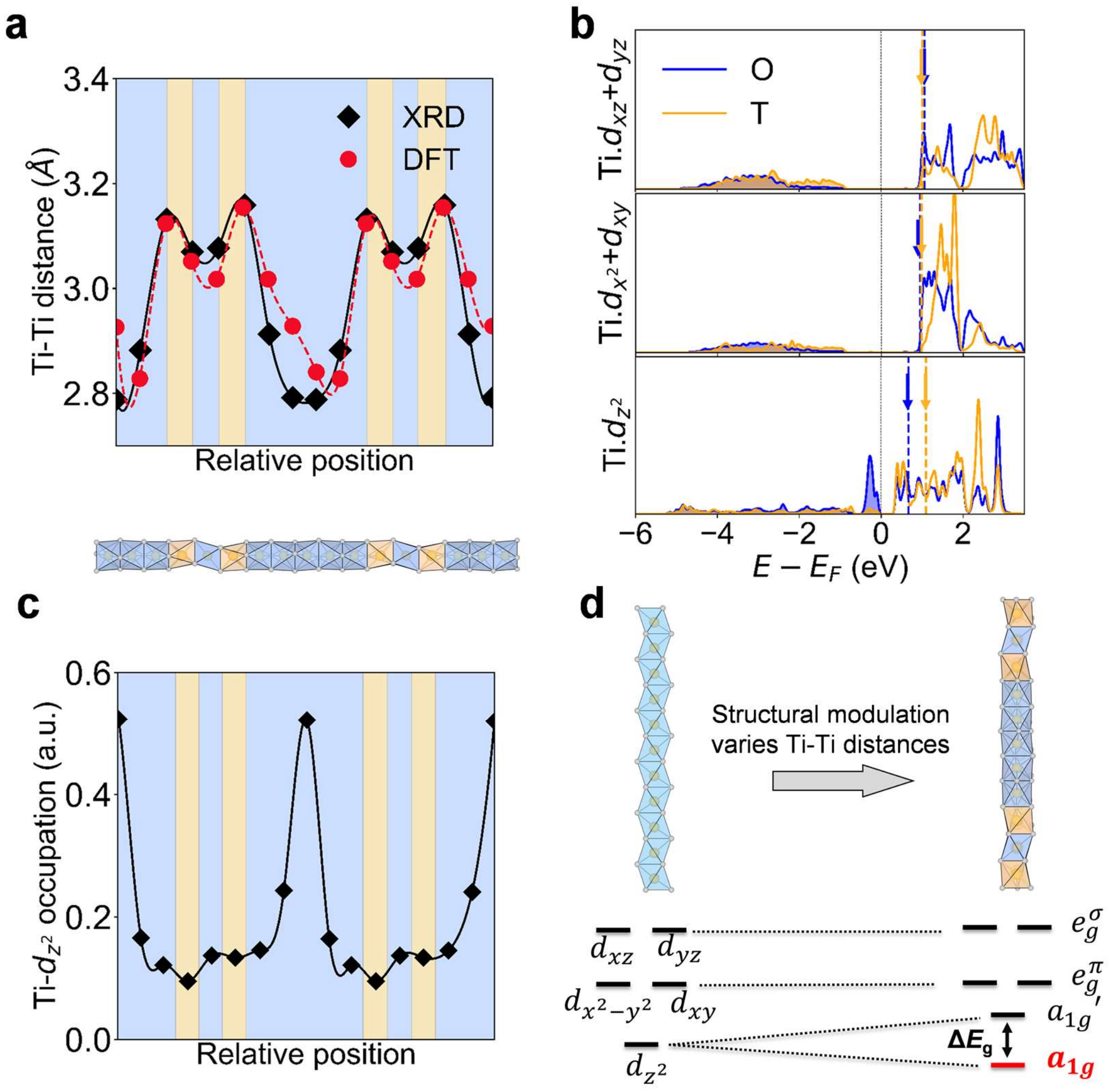

**Supplementary Figure S13. Crystal-field splitting and Ti–Ti distance modulation. a.** The Ti–Ti distance along the *c*-axis as measured directly from XRD refinement compared to their DFT-optimized values. Both XRD and DFT results indicate a larger Ti–Ti distance at the (T−O−T) segments, and smaller Ti–Ti distance at the $(O)_5$ segments. The O and T site positions have been shaded in blue and orange, respectively. An atomic model showing the T and O blocks is shown at the bottom. **b.** PDOS of *d*-states of Ti atoms for the corresponding T and O sites. The center of mass of each state is denoted by blue and orange arrows for T and O sites, respectively. **c.** Ti-$d_{z^2}$ orbital occupation as a function of Ti position along the $TiS_6$ chain. **d.** Schematic of energy levels of *d* states according to PDOS analysis. The Ti–Ti distance modulation arising from structural modulation causes lowers the $d_{z^2}$ ($a_{1g}$) states at the O sites.

**Section 7. STEM-Electron energy loss spectroscopy (EELS)**

In non-stoichiometric $Sr_{9/8}TiS_3$, Ti has an average oxidation state of +3.75, considering that Sr and O have oxidation states of +2 and –2, respectively. However, given the modulated atomic structure of $Sr_{9/8}TiS_3$, DFT calculations suggest selective occupation of $3d_{z^2}$ states of Ti ions within the $(O)_5$ segments. This charge ordering is expected to result in Ti at the $(O)_5$ segments to have an oxidation state lower than the +4 that Ti in the (T−O−T) segments have. To map the local electronic changes in modulated $Sr_{9/8}TiS_3$, we performed STEM-EELS experiments and probed the variation in Ti-$L_{2,3}$ core-loss edge with atomic resolution. The fine structure in a core-loss edge is sensitive to the density of unoccupied electronic states [S10, S11]. The Ti-$L_{2,3}$ edge shows two major peaks: $L_2$ and $L_3$, which arise from the transitions from the inner $2p_{1/2}$ and $2p_{3/2}$ core levels to unoccupied $3d$ states [S12]. The intensity ratio ($L_3/L_2$) has been demonstrated to be sensitive to electron occupation, which is an indicator of the change in oxidation state [S13]. We acquired the core-loss EEL spectra at the Ti-$L_{2,3}$ edge using line-scans that spanned across the (T−O−T) and $(O)_5$ segments and tracked changes in the ($L_3/L_2$) ratio to test the presence of charge ordering.

EELS spectra with atomic-scale spatial resolution were acquired using the Nion UltraSTEM100 operating at 100 kV and equipped with a Gatan Enfina EEL spectrometer. We used a dispersion of 0.1 eV/channel to give a measured energy resolution of 0.4 eV (full width at half maximum of the zero-loss peak), as shown in Fig. S14a. Core-loss EELS data were collected with pixel dwell times of 1 s and a collection semi-angle of 48 mrad. A power-law was used to model the background signal prior to the $L_{2,3}$ edge.

EELS line-scans were acquired across the $(O)_2$−(T−O−T)−$(O)_5$−(T−O−T) segments that are highlighted in the HAADF image in Fig. S14b. A HAADF intensity profile, shown in Fig. S14b, was collected simultaneously with the EELS acquisition to identify the polyhedral units. According to the structural features in modulated $Sr_{9/8}TiS_3$, the atomic columns within the (T−O−T) segment, which has Sr and Ti columns overlapping in the viewing projection, show higher HAADF intensity. By integrating the EEL spectra from the selected region, we compare the Ti-$L_{2,3}$ edge from within the (T−O−T) and $(O)_5$ segments in Fig. S14c. The peak shapes of the Ti-$L_{2,3}$ edge depend on the filling of Ti-$d$ states, and are characteristic of the Ti oxidation state [S13]. Compared to the (T−O−T) segments, the $L_3$ peak — that is indicated by the black arrow in Fig. S14c — has a lower intensity for the $(O)_5$ segment (The intensity of the $L_2$ peak from the two segments has been normalized). Thus, the $L_3/L_2$ ratio within the $(O)_5$ segment is suppressed, which indicates a reduction in the Ti oxidation state.

To relate the variation of the Ti-$L_{2,3}$ edge to the spatial configuration of $TiS_6$ polyhedra, we performed MLLS fitting to decompose the experimental spectra using Ti-$L_{2,3}$ edges from the central O unit of the $(O)_5$ segment and T unit as references:

$$I(E) = AE^{-k} + a_O I_O(E) + b_T I_T(E),$$

where $I(E)$ is total spectra signal, $AE^{-k}$ is the background described using a power-law function, $a_O$ and $b_T$ are weighting factors of Ti-$L_{2,3}$ edges from the central O unit at the $(O)_5$ segment and the T unit, respectively.

In Fig. S14d, the resulting MLLS fitting coefficients are shown as a function of position along the (O)$_2$−(T−O−T)−(O)$_5$−(T−O−T) segment. The oscillation of MLLS fitting coefficients for different Ti-$L_{2,3}$ edges is evidently correlated with the periodicity of the structural modulations. The Ti-$L_{2,3}$ edges from the O unit show major contribution in the (O)$_5$ segments while Ti-$L_{2,3}$ edges from the T unit significantly contributes to the (T−O−T) segments. These results confirm our DFT predictions that the modulated $Sr_{9/8}TiS_3$ lattice has selectively occupied $d_{z^2}$ states localized at (O)$_5$ sites.

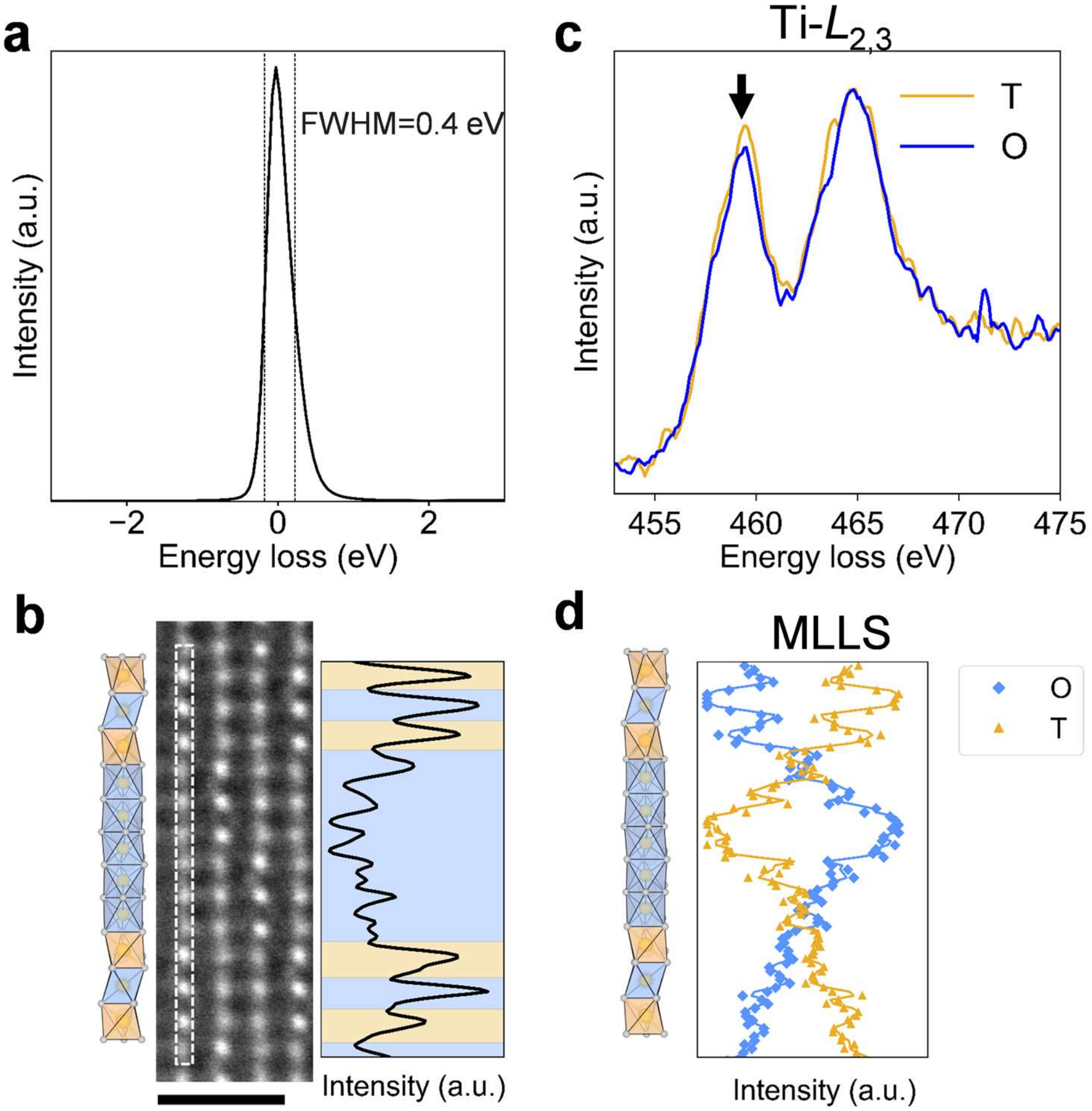

**Supplementary Figure S14. Ti-$L_{2,3}$ EELS. a.** The zero-energy loss peak of EEL spectrum with a full-width at half maximum (FWHM) of 0.4 eV. **b.** HAADF-STEM image showing the region used for EELS acquisition highlighted with the dashed white box. Scale bar is 1 nm. The left-side panel shows the atomic structure with the modulations. The right-side panel shows the HAADF intensity profile collected simultaneously with the EEL spectra. The octahedral sites (O) and pseudo-trigonal prismatic (T) units are shaded in blue and orange, respectively. **c.** Comparison between Ti-$L_{2,3}$ edges extracted from (T−O−T) and (O)$_5$ segments. The $L_3$ peak is labelled with a black arrow. The $L_3/L_2$ ratio of the Ti from the center of the octahedral (O)$_5$ unit is smaller compared to the Ti from the T-site of the (T−O−T) segment indicating a reduced oxidation state of the former. **d**. MLLS intensity profiles showing the spatial variation of the fraction of the two reference Ti-$L_{2,3}$ edges shown in **c.** The spatial variation in the two profiles shows a clear correlation with (T−O−T)−(O)$_5$ periodicity schematically shown in the atomic structure on the left side.

**Section 8. Theoretical calculation of optical properties**

In this study, given the fact that our optical measurements were performed at room temperature on a bulk single crystal, excitonic effects in $Sr_{9/8}TiS_3$ can be neglected. According to the Wannier–Mott model [S14], the exciton binding energy can be calculated as:

$$E_{\mathrm{bind}}=\frac{m_e m_h}{(m_e+m_h)m_0\varepsilon_{\mathrm{stat}}^2}R_{\mathrm{H}},$$

where $m_e$ and $m_h$ are the electron and hole effective masses, respectively, $m_0$ is the free electron mass, $R_H$ is the Rydberg constant (13.6 eV), and $\varepsilon_{\mathrm{stat}}$ is the static dielectric constant calculated as a sum of electronic contribution and ionic contribution. $m_e$ and $m_h$ were obtained by fitting parabola to the band structures at conduction band minimum (CBM) along T-Γ direction and valence band maximum (VBM) along Γ-$S_0$ direction. From our band structure calculations in Fig. 3(b), these values are $0.28m_0$ and $5.47m_0$, respectively. For $Sr_{9/8}TiS_3$, the electronic contribution to $\varepsilon_{\mathrm{stat}}$ along *ab*-plane and *c*-axis are calculated to be 8.3 and 20.9, respectively, at energy/frequencies below the band gap, as shown in Figure 3(e, g). The ionic contributions to $\varepsilon_{\mathrm{stat}}$ were computed by means of density functional perturbation theory (DFPT), which comes to 11 along *ab*-plane and 58 along *c*-axis. To reduce the cost of DFPT calculation for the large size $Sr_{9/8}TiS_3$, we have reduced the density of *k*-point mesh to 2×2×2 and total number of band (NBANDS) to 400 in calculation settings. The total calculated $\varepsilon_{\mathrm{stat}}$ of $Sr_{9/8}TiS_3$ is 19.3 along the *ab*-plane and 78.9 along the *c*-axis. Using these values, we obtain $E_{\mathrm{bind}} \approx$ 9.7 meV along the *ab*-plane, and ~0.6 meV along the *c*-axis. These values are much smaller than the thermal energy (~27 meV) at room temperature, indicating that excitons can easily decompose under the current experimental conditions. To observe any excitonic effect in this material, one would like to perform low-temperature measurements.

Therefore, we assumed the independent particle picture for dielectric function calculations. And the frequency-dependent dielectric function along the different crystal axes were calculated by DFT methods using the VASP code with the LOPTICS tag, which follows the formulation proposed by Gajdoš et al. [S15]

Convergence tests for the number of empty conduction band states were performed for the calculation of the frequency-dependent dielectric function. As shown in Figure S15, with more empty conduction bands added (from 60 eV to 82 eV), we do not observe any significant changes in the calculated dielectric functions. Especially, the dielectric components within *ab*-plane show no discernible changes upon adding more empty states. Dielectric components along the *c*-axis show a slight change on adding more empty states, but it would not affect our conclusion of colossal anisotropic in the dielectric properties. Considering the calculation efficiency and accuracy, we computed the frequency spectrum with transition energies up to 60 eV by setting the total number of energy bands (NBANDS) to be 2.5 times as many as number of valence bands.

To resolve the possible energy transitions, we set the number of grid points on which the density of states (NEDOS tag in VASP) is evaluated to 8000. As indicated in Figure S16, the value of complex shift (CSHIFT tag in VASP) for the Kramers-Kronig transformation was set to 0.05 to achieve the best agreement with the experimental results.

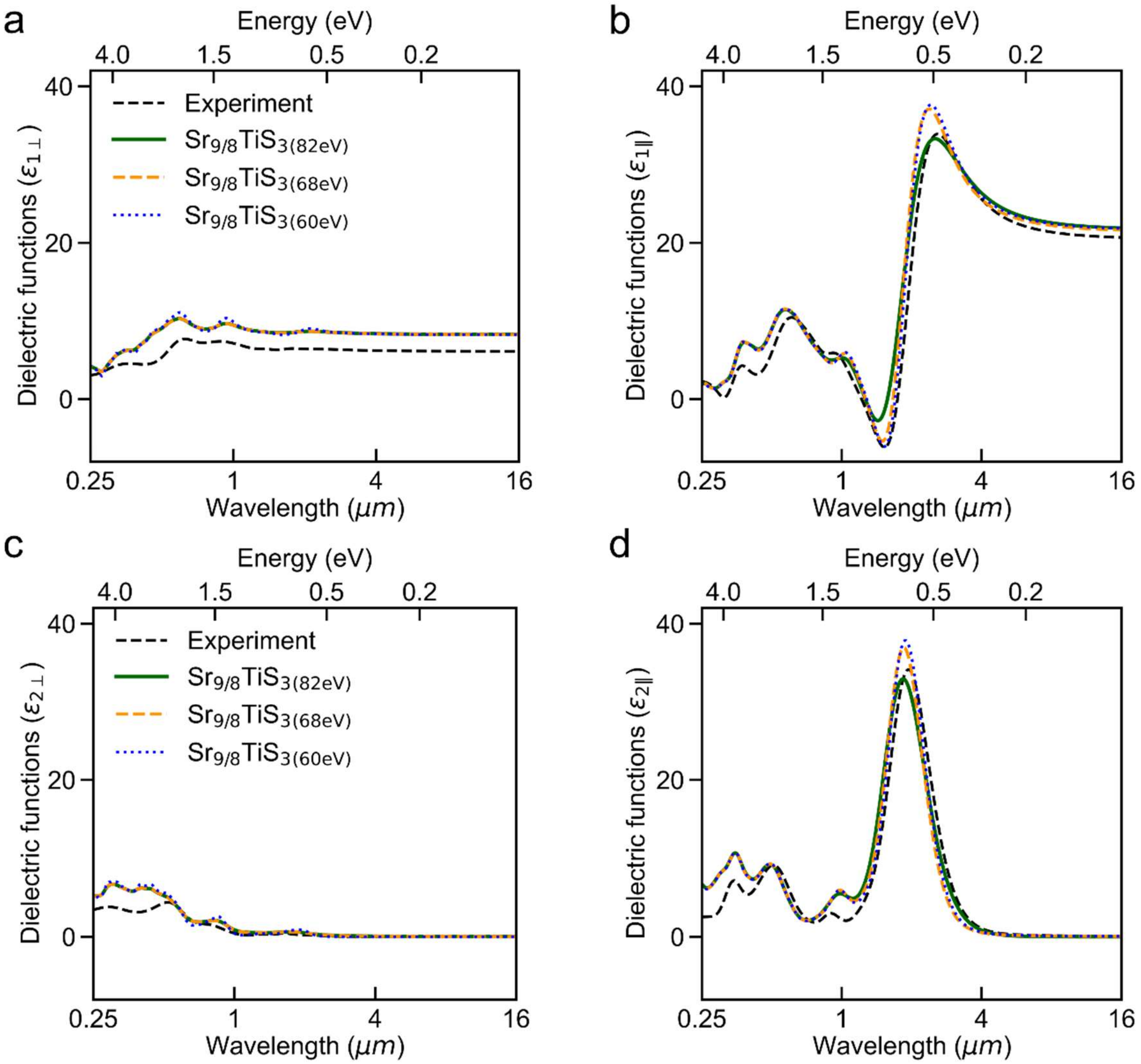

**Supplementary Figure S15. The dependence of the dielectric function with the number of empty bands included in the dielectric function calculations.** Upon adding more empty states (from 60 to 82 eV), dielectric components along the *ab*-plane show no discernible changes for its real part ($\varepsilon_{1\perp}$) shown in **a** and imaginary part ($\varepsilon_{2\perp}$) shown in **c**. For dielectric components along the *c*-axis, real part ($\varepsilon_{1\parallel}$) in **b** and imaginary part ($\varepsilon_{2\parallel}$) in **d** only show a slight change as add more empty states.

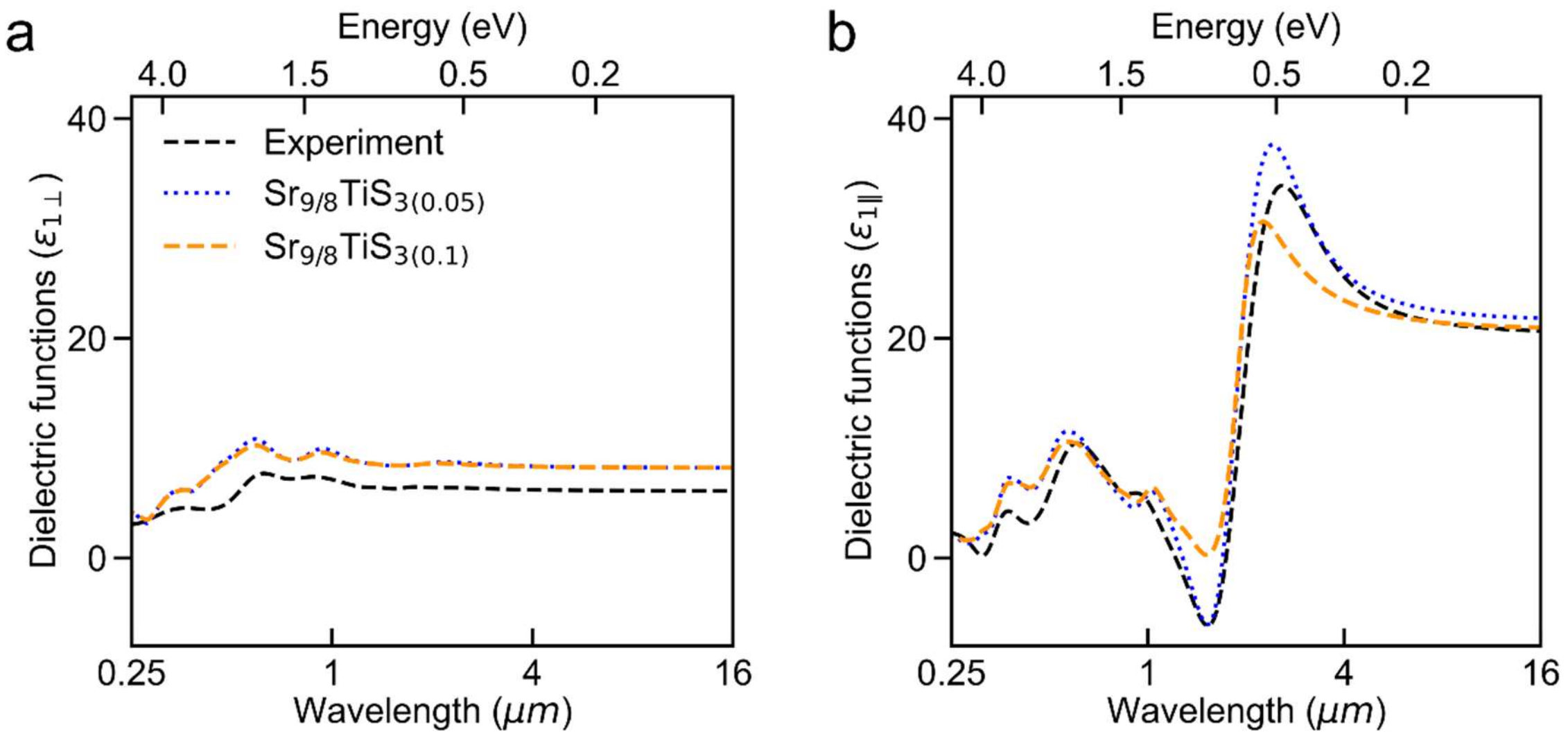

**Supplementary Figure S16.** The effect of the complex shift parameter $\eta$ (defined using the CSHIFT tag in VASP) on the real part of the dielectric function calculated using the Kramers-Kronig transformation. CSHIFT equal to 0.05 was found to give the best agreement with the experimental results.

The band-to-band transition strength was analyzed by calculating the squares of the dipole transition matrix elements along different $k$-point paths. As plotted in Figure S17b, we find the transitions between the band edges have the largest contribution to the enhanced $\varepsilon_2$ at low frequencies.

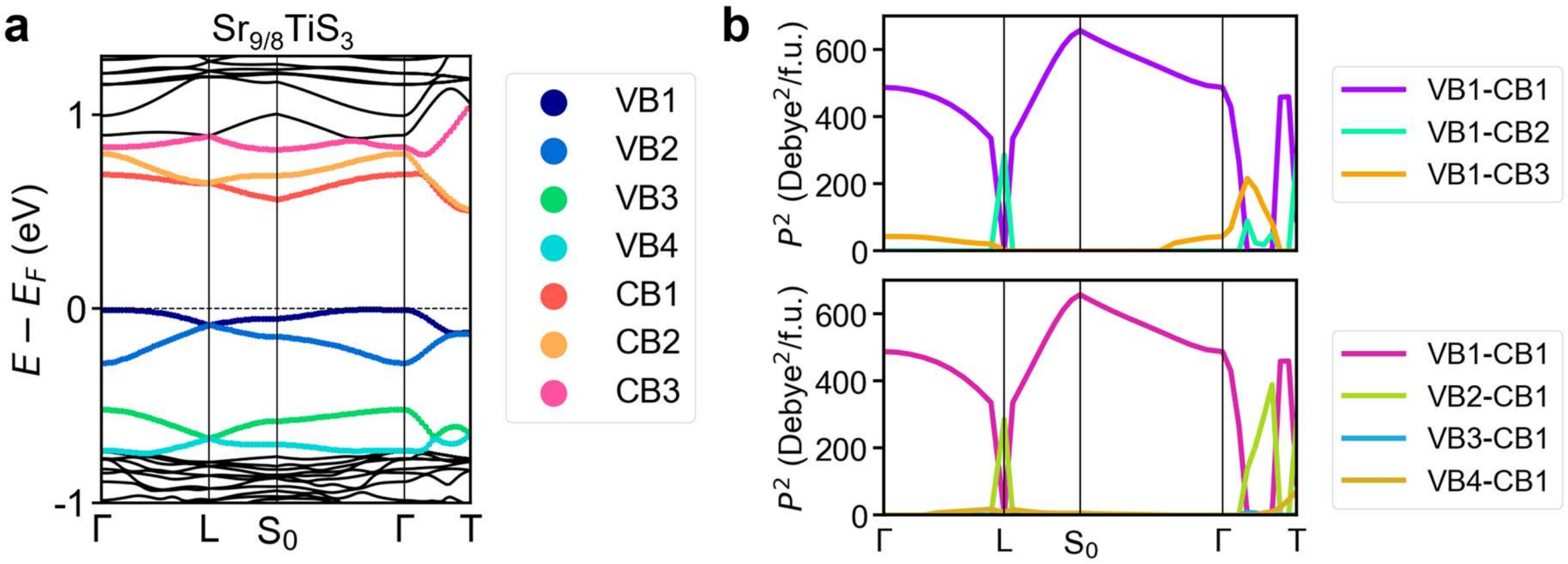

**Supplementary Figure S17 a.** PBE+U calculated band structure of $Sr_{9/8}TiS_3$. The top 4 valence bands and the 3 lowest conduction bands have been indexed with different colors. **b.** The squares of the dipole transition matrix elements for different energy transitions from valence bands to conduction bands. The upper panel compares the transition strength from the top valence band (VB1) to three lowest conduction bands (CB1, CB2, CB3). In the bottom panel, the transition strength from different valence bands (VB1, VB2, VB3, VB4) to the lowest conduction band (CB1) has been compared. From these comparisons, the lowest interband transition involving VB1 – CB1 has a dominant effect over other higher energy transitions.

Squares of the dipole transition matrix elements along the $ab$-plane and the $c$-axis are plotted for comparison, as shown in Figure S18. The band edges transitions (VB1-CB1 and VB2-CB1) for the selectively occupied $3d_{z^2}$ bands show the strongest transition strength along $c$-axis ($z$-direction in the Figure S18), but they are much weaker along the *ab*-plane (*x*-, *y*-directions in the figure). The bandgap for lowest energy transition along two directions can be found from an enlarged spectrum of imaginary part of dielectric functions around band edges transitions, where $\varepsilon_{2\parallel}(\omega)$ corresponds to dipole transition along the *c*-axis, while $\varepsilon_{2\perp}(\omega)$ corresponds to dipole transition along *ab*-plane. As shown in Figure S18(c-d), the lowest transition energy for *ab*-plane electric dipole and *c*-axis electric dipole have a similar transition energy onset of ~0.4 eV between the top valence band and the bottom conduction band.

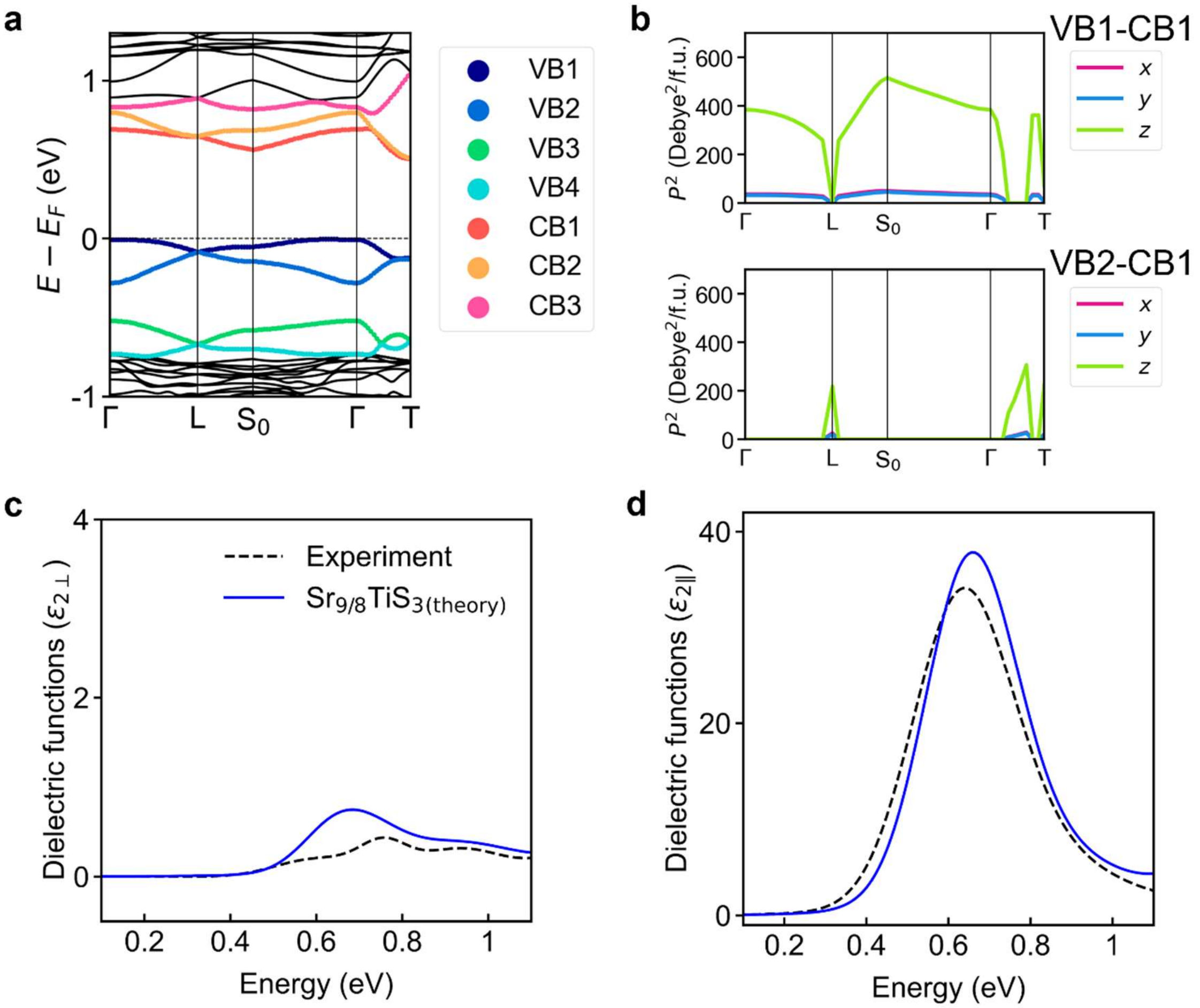

**Supplementary Figure S18**. **a.** PBE+*U* calculated band structure of $Sr_{9/8}TiS_3$. The top 4 valence bands and the bottom 3 conduction bands have been indexed with different colors. **b.** The squares of the dipole transition matrix elements for band edge transitions along the *c*-axis (*z*-direction; top panel) and the *ab*-plane (*x*-, *y*-directions; bottom panel). **c-d.** The imaginary part of the dielectric function for the lowest interband transitions. For the transition from top valence band to lowest conduction band, the transition dipole along *ab*-plane ($\varepsilon_{2\perp}$) in **c** shows a similar transition energy onset of ~0.4 eV with the transition dipole along *c*-axis ($\varepsilon_{2\parallel}$) in **d**.

Based on the computed frequency-dependent dielectric function, optical properties, such as the real part of the refractive index, $n(\omega)$, and the extinction coefficient, $\kappa(\omega)$, can be obtained in terms of the real part, $\varepsilon_1(\omega)$, and imaginary part of the dielectric function, $\varepsilon_2(\omega)$:

$$\mathrm{n}(\omega) = \frac{\sqrt{2}}{2}\left[\sqrt{\varepsilon_1(\omega)^2 + \varepsilon_2(\omega)^2} + \varepsilon_1(\omega)\right]^{1/2},$$

$$k(\omega) = \frac{\sqrt{2}}{2}\left[\sqrt{\varepsilon_1(\omega)^2 + \varepsilon_2(\omega)^2} - \varepsilon_1(\omega)\right]^{1/2}$$

The optical anisotropy has been quantified in terms of birefringence and dichroism. In this work, we consider the *c*-axis as the optical axis. The birefringence was calculated by taking the difference between refractive indices for the polarization along the *c*-axis and perpendicular to the *c*-axis. The dichroism was calculated in the same manner using the extinction coefficients.

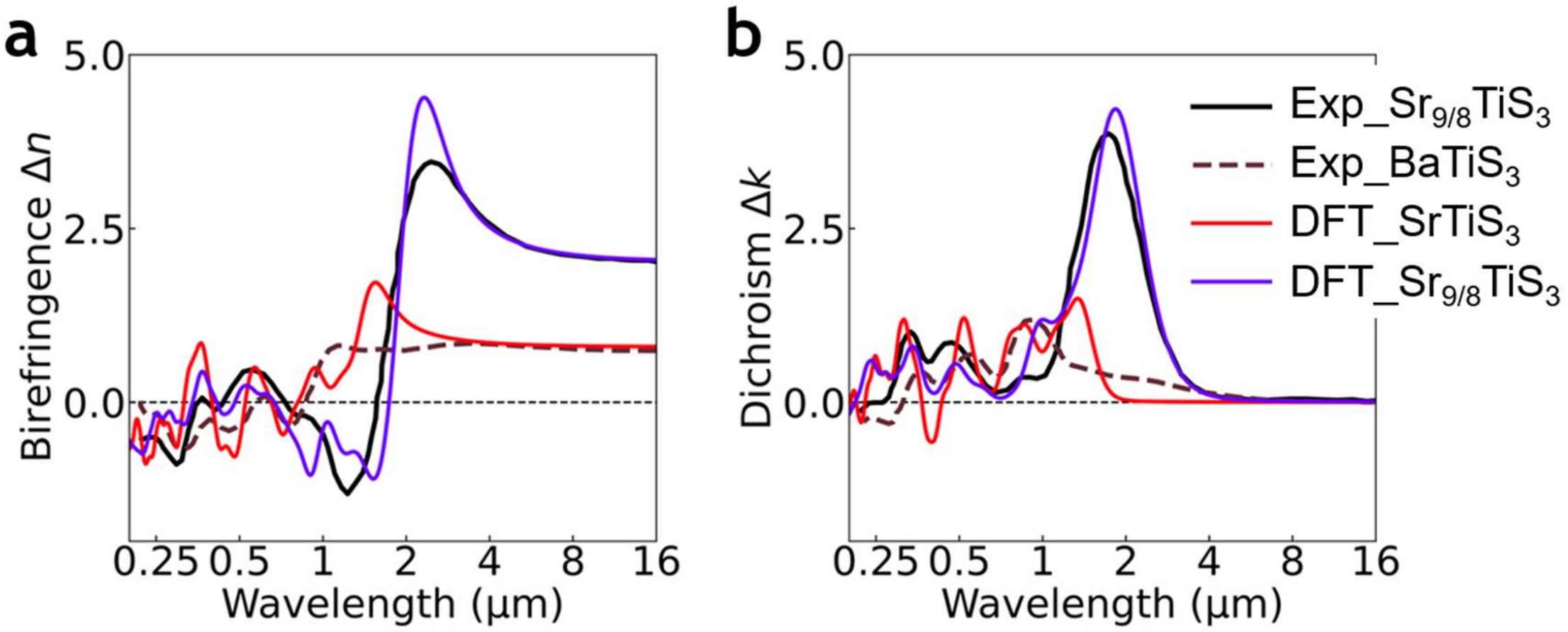

**Supplementary Figure S19. Optical anisotropy. a-b.** Birefringence and dichroism comparison between experimental measurements for $Sr_{9/8}TiS_3$ (in black solid), $BaTiS_3$ (in black dash) and DFT calculations for $Sr_{9/8}TiS_3$ (in blue solid) and $SrTiS_3$ (in red solid). All DFT calculations were performed at the GGA+*U* level (*U* = 3 eV for Ti atoms). The hypothetical stoichiometric $SrTiS_3$ has a similar optical anisotropy as $BaTiS_3$.

We account for strong correlation effects within the localized *d* states of Ti by using the GGA + Hubbard *U* method [S16]. The Hubbard *U* value was selected to be 3 eV as it was found to best match the experimental results, as shown in Fig. S20.

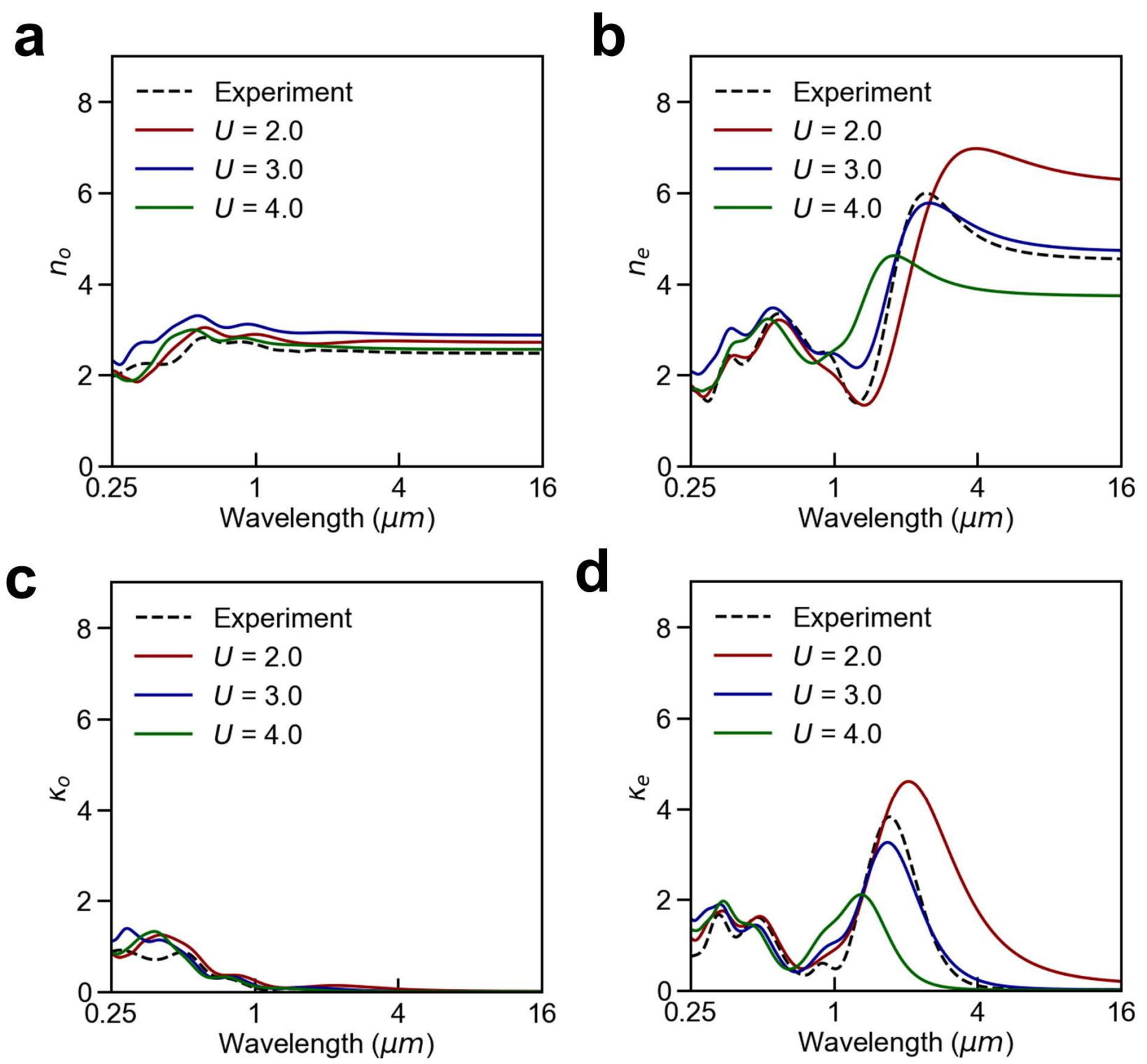

**Supplementary Figure S20. The effect of Hubbard *U* on calculated optical properties of $Sr_{9/8}TiS_3$. a-b.** Real part of complex refractive index along the ordinary axis ($n_o$) and the extraordinary axis ($n_o$), **c-d.** Imaginary part of the complex refractive index along the ordinary axis ($\kappa_o$), and the extraordinary axis ($\kappa_e$).

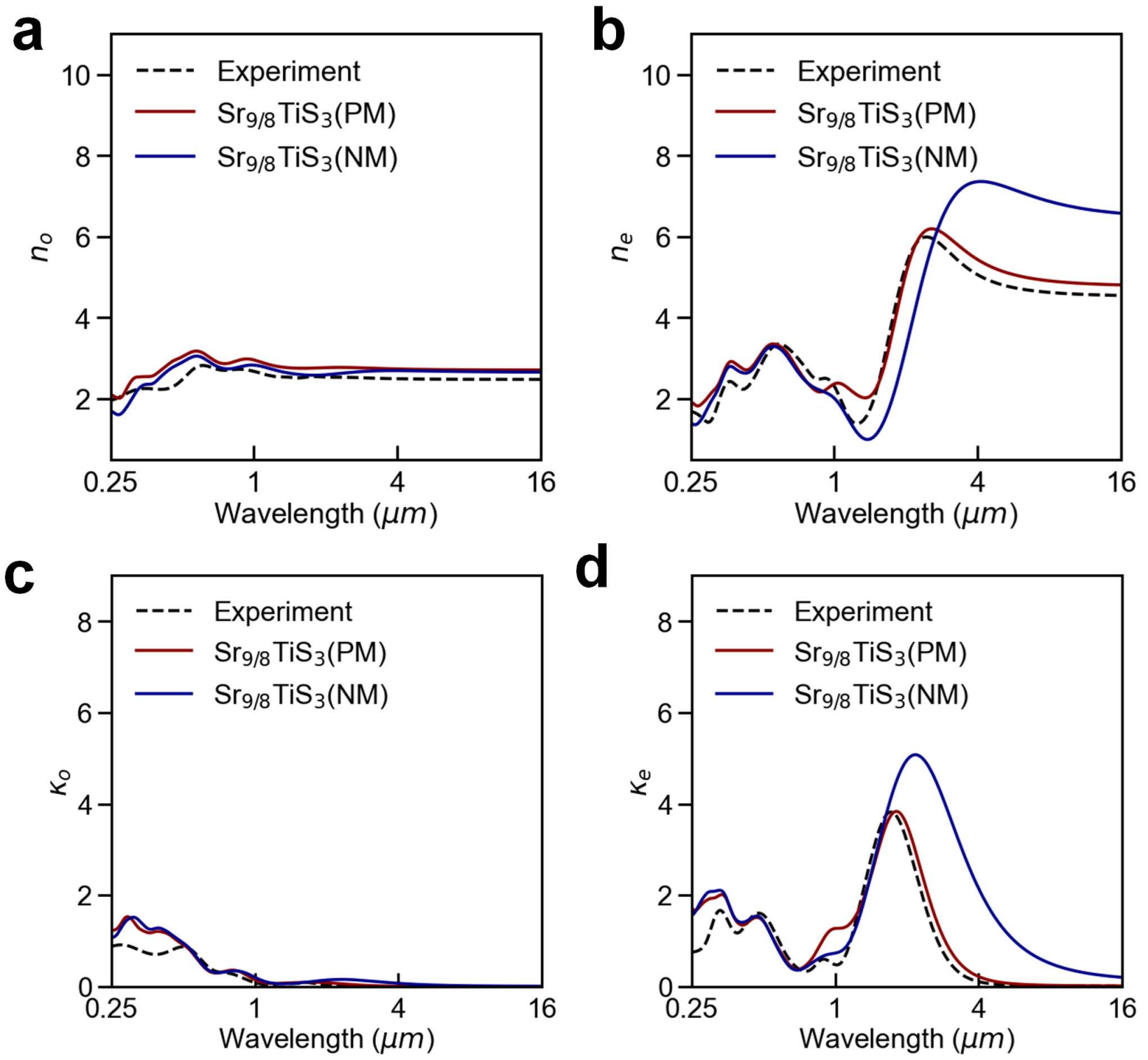

**Supplementary Figure S21. Calculated optical anisotropy of $Sr_{9/8}TiS_3$ with different magnetic orderings. a.** real part of the refractive indices along ordinary axis ($n_o$), **b** real part of the refractive indices along extraordinary axis ($n_e$), **c.** extinction coefficient along ordinary axis ($\kappa_o$), **d.** extinction coefficient along extraordinary axis ($\kappa_e$).

## Section 9. Initial demonstration of a $Sr_{9/8}TiS_3$ wave plate

To demonstrate the use of $Sr_{9/8}TiS_3$ plate as a wave plate, we performed a first-order experiment with light propagating through two crossed polarizers, with a $Sr_{9/8}TiS_3$ plate with thickness of 3.2 μm in between. When the sample is rotated, the transmission varies from ~0 to ~90% due to polarization rotation within the birefringent material.

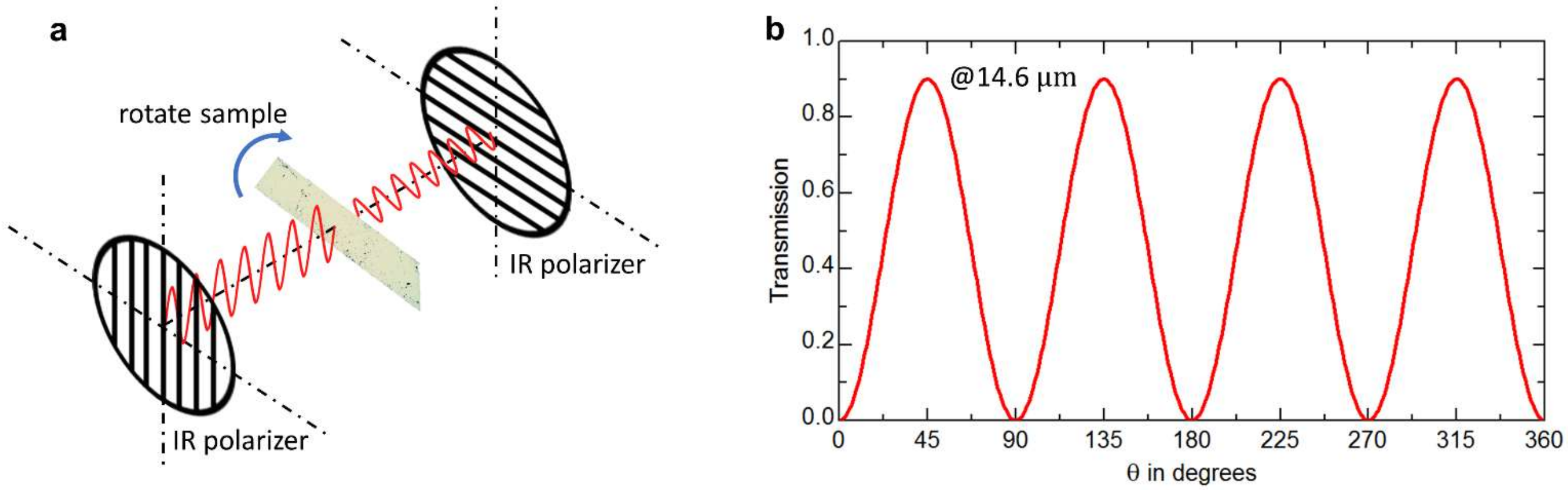

**Supplementary Figure S22. a.** Schematics of the measurement with two crossed IR polarizers and the $Sr_{9/8}TiS_3$ sample in between. **b.** Transmission at 14.6 μm as a function of rotation angle of the sample.

## Additional References